\documentclass[lettersize,journal]{IEEEtran}
\usepackage{amsmath,amsfonts}
\usepackage{algorithmic}
\usepackage{bm}
\usepackage{algorithm}
\usepackage{array}
\usepackage{graphicx}
\usepackage{subfigure}
\usepackage{textcomp}
\usepackage{stfloats}
\usepackage{url}
\usepackage{verbatim}
\usepackage{pifont}
\usepackage{cite}
\usepackage{tablefootnote}
\usepackage{subfigure}
\usepackage{makecell}
\usepackage{multirow}
\usepackage{booktabs}
\usepackage{float} 

\begin{document}
\title{GenMetaLoc: Learning to Learn Environment-Aware Fingerprint Generation for Sample Efficient  Wireless Localization

\author{Jun Gao, Feng Yin, Wenzhong Yan, Qinglei Kong, Lexi Xu, Shuguang Cui}

\thanks{J. Gao is with the Future Network of Intelligence Institute (FNii), and also with the School of Science and Engineering (SSE), The Chinese University of Hong Kong, Shenzhen, China (e-mail: jungao@link.cuhk.edu.cn).}
\thanks{F. Yin is with the School of Science and Engineering (SSE), The Chinese University of Hong Kong, Shenzhen, China (e-mail: yinfeng@cuhk.edu.cn).}
\thanks{W. Yan is with the School of Science and Engineering (SSE), The Chinese University of Hong Kong, Shenzhen, China (e-mail: wenzhongyan@link.cuhk.edu.cn).}
\thanks{Q. Kong is with the Institute of Space Science and Applied Technology, Harbin Institute of Technology (Shenzhen), and also with the Guangdong Provincial Key Laboratory of Future Networks of Intelligence, Shenzhen, China (e-mail: kql8904@163.com).}
\thanks{L. Xu is with the Research Institute, China United Network Communications Corporation, Beijing 100048, China (e-mail: davidlexi@hotmail.com)}
\thanks{S. Cui is with the School of Science and Engineering (SSE), the Future Network of Intelligence Institute (FNii), and the Guangdong Provincial Key Laboratory of Future Networks of Intelligence, The Chinese University of Hong Kong; he is also affiliated with Peng Cheng Laboratory, Shenzhen, China (e-mail: shuguangcui@cuhk.edu.cn).}}

\maketitle

\begin{abstract}
Existing fingerprinting-based localization methods often require extensive data collection and struggle to generalize to new environments. In contrast to previous environment-unknown MetaLoc~\cite{gao2022metaloc,10447939,10274764}, we propose GenMetaLoc in this paper, which first introduces meta-learning to enable the generation of dense fingerprint databases from an environment-aware perspective. In the model aspect, the learning-to-learn mechanism accelerates the fingerprint generation process by facilitating rapid adaptation to new environments with minimal data. Additionally, we incorporate 3D point cloud data from the first Fresnel zone between the transmitter and receiver, which describes the obstacles distribution in the environment and serves as a condition to guide the diffusion model in generating more accurate fingerprints. In the data processing aspect, unlike most studies that focus solely on
channel state information (CSI) amplitude or phase, we present a
comprehensive processing that addresses both, correcting errors from WiFi hardware limitations such as amplitude discrepancies
and frequency offsets. For the data collection platform, we develop an uplink wireless localization system that leverages the sensing capabilities of existing commercial WiFi devices and mobile phones, thus reducing the need for additional deployment costs. Experimental results on real datasets show that our framework outperforms baseline methods.
\end{abstract}


\section{Introduction}
With the rapid development of the Internet of Things (IoT), precise localization and environmental sensing technologies are crucial for advancements in smart homes~\cite{chen2022fidora,xing2024constructing}, healthcare~\cite{wang2020csi}, and smart city management~\cite{10230036}.  While global navigation satellite systems (GNSS) are  extensively applied for outdoor localization, they often perform poor in indoor environment due to signal blockage by buildings~\cite{7275552}. Although the geometric localization method~\cite{10297296} for indoor environments have been widely explored, it often faces accuracy issues in non-line-of-sight (NLOS) scenarios.   As an alternative, fingerprinting-based localization offers a practical solution for indoor areas by utilizing the spatial distribution of wireless signal characteristics, mapping signal features as fingerprints to corresponding locations. Recently, improvements in technologies have introduced various wireless systems, such as WiFi~\cite{li2021transloc}, ultra wideband (UWB)~\cite{feng2020kalman}, and 6G~\cite{9737357}. These systems provide diverse features for fingerprinting-based localization, such as received signal strength indicator (RSSI)~\cite{yin2017received}, and channel state information (CSI)~\cite{li2021transloc}. In line with these advancements, a new task group (TG) IEEE 802.11bf has been established to update the WLAN standard, addressing advanced sensing requirements while minimizing the impact on traditional communication functions~\cite{du2024overview}, which aligns with the ongoing evolution towards integrated sensing and communication (ISAC).

However, achieving high localization accuracy presents some challenges due to the complexity of indoor wireless propagation environments.  Specifically, environmental changes, such as moving objects or people, can significantly affect the multipath components, leading to fluctuations in the received signals. Consequently, the original fingerprint database becomes outdated, necessitating the collection of new data to describe the changed environment. On the other hand, increasing the density of the fingerprint database can further enhance localization accuracy~\cite{4536547}, but it comes at the cost of substantial human and material resources. To address these challenges, various machine learning techniques have been proposed, such as data augmentation~\cite{sinha2019data}, semi-supervised learning~\cite{shrivastava2017learning}, and informed machine learning~\cite{von2021informed}. Despite these efforts, most studies focus on specific environments~\cite{bahl2000radar,yin2017received,torres2014ujiindoorloc}, making it difficult to develop models that are effective across different indoor environments. As a result, a model that works well in one environment may not perform well in another. This highlights the need for a machine learning model capable of learning essential channel features and being broadly applicable across various indoor environments, as emphasized in a recent 6G white paper~\cite{ali20206g}. 

To meet the above challenges, our prior works first introduced meta-learning into wireless localization and proposed MetaLoc from both deterministic~\cite{gao2022metaloc,10447939} and Bayesian perspectives~\cite{10274764}. Specifically, the underlying localization model is implemented using a deep neural network (DNN), where the input is the wireless fingerprint and the output is the corresponding location. We train the meta-parameters using historical data collected from a variety of well-calibrated indoor environments. These meta-parameters are then used to initialize the neural network, enabling rapid adaptation to new environments with minimal data samples. Similar to most existing works~\cite{xing2022integrated,zhu2022intelligent,li2021transloc,CRISLoc}, our previous approach assumes the location of WiFi access points (APs) and environmental information are unknown. However, the wireless fingerprint is strongly influenced by the propagation characteristics in the environment, such as the distribution of obstacles. These environmental factors directly affect multipath propagation, which in turn impacts the accuracy of localization~\cite{8559912}. Therefore, uncovering  environmental context, such as the distribution of obstacles in the signal path, is essential for improving the precision and robustness of the localization system,  an aspect that remains relatively overlooked.

\subsection{Related Works}
The fingerprinting-based localization method leverages the spatial distribution of signal characteristics, mapping signal features as fingerprints to corresponding locations for localization. It typically consists of two main stages: the offline stage and the online stage. In the offline stage, wireless signals are collected at designated RPs to create a fingerprint database. In the online stage, signals from a test point (TP) are compared with this database to determine the location. Fingerprinting-based localization is generally classified into deterministic and probabilistic approaches.
In deterministic localization, the target’s location is estimated by comparing received signal features with the pre-stored fingerprints, and $K$-nearest neighbors (KNN) algorithm is the most widely used method as introduced in RADAR~\cite{bahl2000radar}. Similarity measures for this approach usually include Euclidean distance \cite{wu2017gain}, cosine similarity \cite{he2014sectjunction}, and Tanimoto similarity \cite{jiang2012ariel}. However, deterministic methods are sensitive to both signal noise and the curse of dimensionality, which can degrade localization accuracy, particularly in high-dimensional multi-carrier CSI of large-scale antenna systems. On the other hand, probabilistic localization employs statistical models to estimate the target’s location by comparing received signal measurements with a fingerprint database. Methods such as Horus~\cite{youssef2005horus} and Bayesian networks \cite{jin2020bayesian} model signal distributions and use probabilistic inference to make predictions.  DeepFi~\cite{wang2015deepfi} further improves computational efficiency by combining probabilistic models with greedy learning algorithms. While probabilistic methods typically require fewer computational resources, they are more complex to implement in dynamic environments, as they rely on precise position-related data.

In recent years, machine learning has played a significant role in the field of fingerprinting-based localization. Ghzali \textit{et al.} attempted to address the difficulties associated with indoor localization by approaching it as a regression problem based on RSSI gathered in a real office space. Their proposed solution relied on a neural network with random initialization~\cite{ghozali2019indoor}. ConFi~\cite{chen2017confi} was the first work to  use the convolutional neural networks (CNNs) for learning CSI images at RP, opening up new possibilities for indoor localization. Hsieh \textit{et al.} attempted to solve the indoor localization challenge by utilizing the RSSI and CSI data as a classification problem. They evaluated various neural network architectures to find the best fit for accurately estimating the location of an object within a specific room~\cite{hsieh2019deep}. Although machine learning has shown promise in fingerprinting-based localization, it faces two main challenges. Firstly, the dynamic nature of wireless environments introduces adaptability issues, and models may struggle to maintain accuracy in the changing signal conditions. Secondly, certain machine learning methods are highly data-hungry, requiring substantial human effort for data collection.

To address the first challenge, recent works are being explored on both the model and data sides. On the model side, domain adaptation techniques such as transfer learning are widely used, where the source domain represents the original environment and the target domain is a new or unseen environment. For example, TransLoc~\cite{li2021transloc} employs transfer learning to identify cross-domain mappings and create a homogeneous feature space containing discriminative information from multiple domains. CRISLoc\cite{CRISLoc} uses transfer learning to reconstruct a high-dimensional CSI fingerprint database with outdated fingerprints and a few new measurements. Fidora\cite{chen2022fidora} adapts to new data using a variational autoencoder and a joint classification-reconstruction structure. ILCL\cite{zhu2022intelligent} employs incremental learning and expands neural nodes to reduce training time. On the data side, ViVi\cite{wu2017gain} reduces uncertainty in RSSI fingerprints by exploiting spatial gradients across multiple locations. CiFi\cite{cifi} enhances the stability of CSI fingerprints by using phase differences between antenna pairs instead of raw measurements. DFPS~\cite{fang2015novel} combines raw RSSI with differences between AP pairs to improve adaptability against heterogeneous hardware.

However, the works mentioned above primarily focus on wireless signals, often overlooking the integration of environmental information. Recently, some radio map methods have incorporated environmental data to generate signal strength at each point, providing valuable insights for fingerprint-based localization. For example, in~\cite{9354041}, the authors framed radio map construction as an image-to-image prediction task, where environmental data and transmitter locations were represented as two images. Using CNNs, the proposed RadioUNet predicted signal strength based on a ray-tracing simulation dataset with various environments. Meanwhile, in~\cite{yang2024orchloc}, environmental obstacles were modeled using manually drawn geometries, and their distribution was determined through point sampling, allowing them to obtain environmental information for subsequent outdoor localization. While these methods incorporate environmental data, they rely on 2D projection of the scenarios~\cite{9354041,9918126} or manually drawn approach~\cite{yang2024orchloc}, losing some details. Recently, point cloud data obtained through LiDAR techniques has been studied in autonomous vehicles~\cite{9307334} and robotics~\cite{zhou2021path}, motivating its integration into localization systems by providing detailed 3D environmental information.

To address the second challenge, we have first introduced meta-learning into wireless localization and proposed MetaLoc~\cite{gao2022metaloc,10447939, 10274764}, yielding promising results under environment-unknown conditions. Recently, the sim-to-real concept~\cite{yu2024learning} has been proposed, generating virtual environments that mimic real-world conditions and using synthesized data for model training to reduce reliance on real-world data, thus minimizing the need for extensive datasets. This approach inspires our work towards a new direction for environment-aware localization, offering the potential for a cost-effective indoor fingerprinting-based method with fewer site surveys.
\subsection{Contributions}
In contrast to our previous MetaLoc without considering environmental information, in this paper we propose GenMetaLoc, an environment-aware localization framework that leverages existing WiFi devices. Point cloud data from the first Fresnel zone between a known transmitter and receiver is used to enhance fingerprint generation. During the meta-training stage, this data is processed through a diffusion model with a U-Net and attention mechanism to learn meta-parameters via a two-loop procedure, establishing a mapping from the environment to wireless fingerprints. In the meta-test stage, when the environment changes, the fingerprint generation model is rapidly adapted using sparse fingerprints, leveraging the meta-parameters learned during meta-training. Overall, the main contributions of this work are fourfold:
\begin{itemize}
\item \textbf{Model Design}:  In contrast to previous environment-unknown
MetaLoc, we first introduce meta-learning to generate dense fingerprint databases from an environment-aware perspective. Specifically, point cloud data from the first Fresnel zone is incorporated as conditioning in the diffusion model, capturing the obstacle distribution in the line-of-sight (LOS) area. To reduce the computational workload in high-dimensional multi-subcarrier data, we utilize an autoencoder to compress the raw data into a lower-dimensional representation. Additionally, we apply a slight Kullback-Leibler (KL)-penalty term to ensure the learned latent space approximates a standard normal distribution, preventing arbitrarily high variance in the latent space.
\item \textbf{Collection Platform}: To validate the proposed framework, we establish a more sophisticated uplink wireless localization platform that utilizes the sensing capabilities of existing commercial WiFi devices. In contrast to MetaLoc that relies on sequential data reception from multiple APs in a downlink system, the system in GenMetaLoc enables the central server to simultaneously control five APs to receive data, making the platform practicality and eliminating the need for specialized data collection equipment, thereby reducing the deployment costs.
\item \textbf{Data Processing}: In contrast to most studies that focus solely on CSI amplitude~\cite{CRISLoc,gao2022metaloc} or phase~\cite{wang2015deepfi,zhu2022intelligent}, our work offers a  comprehensive processing method that addresses both. Specifically, we mitigate errors arising from the hardware limitations of WiFi devices, such as amplitude discrepancies due to automatic gain control (AGC) and frequency offsets caused by unsynchronized clocks. Through these corrections, our method generates cleaner and more accurate CIR data after the inverse fast Fourier transform (IFFT) of CSI, enhancing the subsequent localization process.
\item \textbf{Localization Effectiveness}: A dense fingerprint database can be generated with minimal real-world data in new environments through a learning-to-learn mechanism. By transforming the diffusion model into a data source for generating CIR conditioned on diverse environmental inputs, the proposed GenMetaLoc facilitates the sim-to-real transition. Specifically, the localization model is trained using synthetic CIR data from the diffusion model, and this trained model is subsequently applied to real-world environments, thus reducing the need for extensive field data collection.
\end{itemize}

\section{FINGERPRINTING-BASED LOCALIZATION: SCENARIO,
MODELING, AND DATA PROCESS}
In this section, we first describe the scenario for CIR fingerprinting-based localization in Sec.~\ref{section:scenatio}. Next, we introduce the channel model in Sec. ~\ref{section:model}, and  finally we discuss the wireless data and environmental data preprocessing in Sec.~\ref{section:process} and Sec.~\ref{sec:environmantal information}, respectively.
\subsection{Scenario Description }~\label{section:scenatio}
\begin{figure}
    \centering
    \includegraphics[scale=0.78]{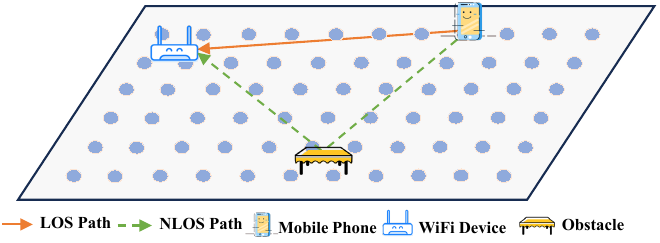}
    \caption{The scenario of the  fingerprinting-based localization in an uplink wireless communication system.}
    \label{fig:area}
\end{figure}
As illustrated in Fig.~\ref{fig:area}, we consider an uplink communication system consisting of $Z$ fixed WiFi APs and a user equipped with a mobile phone. Both the APs and the phone are equipped with omnidirectional antennas. Specifically, we define a survey area of size $A=(v \times w)\mathrm{m}^2$ with $v$ and $w$ representing the length and width, respectively. The area is divided into $G=\left\lceil\frac{v}{\Delta r}\right\rceil \times \left\lceil\frac{w}{\Delta r}\right\rceil$ grid points, and each grid point is denoted as $\left\{\boldsymbol{p}_g=\left(x_g, y_g, h_g\right) \mid g=1, \cdots, G\right\}$, with equal spacing of $\Delta r \ \mathrm{m}$. Here, $\lceil \cdot \rceil$ represents the ceiling function, and $\left\{\boldsymbol{p}_g=\left(x_g, y_g, h_g\right) \mid g=1, \cdots, G\right\}$ refers to the set of locations in a 3D Cartesian coordinate system.

\subsection{Channel Modeling}~\label{section:model}
As illustrated in Fig.~\ref{fig:area}, the communication channel between the user and the APs is composed of a direct LOS path and several indirect NLOS paths including reflections, diffractions, and scattering. Both frequency domain CSI and time domain CIR are essential for capturing the small-scale multipath effects in wireless channel measurements. Specifically, each CSI is represented as:

\begin{equation} H\left(f_k\right)=\left|H\left(f_k\right)\right| e^{j \sin (\angle H(f_k))}, \quad k=1, \ldots, K, \label{eq:csi} \end{equation}

where $\left|H\left(f_k\right)\right|$ denotes the amplitude, and $\angle H(f_k)$ represents the phase of the CSI at the central subcarrier frequency $f_k$. The total number of subcarriers $K$ depends on the channel bandwidth, as specified by the IEEE 802.11ac standards: 64 for 20 MHz, 128 for 40 MHz, and 256 for 80 MHz. Meanwhile, CIR $h(\tau)$ is defined as:
\begin{equation} h(\tau)=\sum_{i=1}^N a_i e^{-j \theta_i} \delta\left(\tau-\tau_i\right), \label{eq:cir} \end{equation}
where the $i$-th term in the CIR corresponds to a propagation path characterized by a time delay $\tau_i$, amplitude attenuation $a_i$, and phase shift $\theta_i$, with $N$ denoting the total number of paths. In summary, CSI and CIR offer complementary views of channel characteristics. While many studies focus on either the CSI amplitude or phase, we  consider both and converse the complete CSI information into CIR via IFFT, thereby preserving more channel details.
\subsection{Wireless Data Preprocessing}~\label{section:process}
Capturing wireless data usually requires expensive specialized equipments such as vector network analyzers (VNA) or software-defined radios (SDR). A more cost-effective alternative is to use the existing commercial WiFi devices, which can capture frequency domain CSI from OFDM-modulated frames with up to 80 MHz bandwidth on some Broadcom WiFi chips~\cite{gringoli2019free}. However, due to limitations in WiFi hardware, this approach faces some challenges. The CSI obtained from commercial WiFi devices may include distortions due to signal superposition during propagation and hardware issues, such as inaccurate sampling frequencies and central frequency shifts~\cite{CRISLoc,8423070}. These errors can significantly affect the derived CIR, altering the multipath characteristics~\cite{8423070}. To mitigate these issues, we will discuss CSI data preprocessing methods including abnormal frame removal, CSI amplitude calibration, and phase fitting in the following sections.
\subsubsection{Abnormal Packet Removal}
\begin{figure}
    \centering
    \includegraphics[scale=0.55]{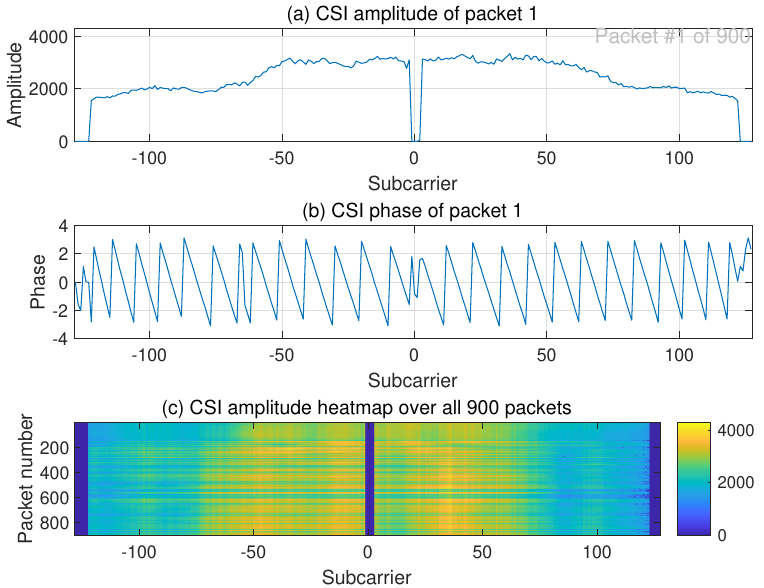}
    \caption{CSI measurements collected from the commercial WiFi device ASUS RT-AC86U in a stationary point: (a) CSI amplitude across subcarriers for the first packet.
(b) CSI phase across subcarriers for the first packet. (c) CSI amplitude heatmap over all 900 collected packets, with lighter areas indicating higher amplitudes.
}
    \label{fig:csi-example}
\end{figure}

Figure~\ref{fig:csi-example} displays CSI measurements obtained from the ASUS RT-AC86U WiFi device at a stationary point. Specifically, Fig.~\ref{fig:csi-example} (a) and (b) show the amplitude and phase of CSI across subcarriers for the first packet, respectively. Figure~\ref{fig:csi-example} (c) illustrates a CSI amplitude heatmap for all 900 collected packets, where lighter areas indicates higher amplitudes. It is noted that guard subcarriers from -128 to -123, -1 to 1, and 123 to 127 may contain random values, so they are typically set to zero to prevent disturbance as recommended in~\cite{nexmon:project}. 

As shown in Fig.~\ref{fig:csi-example} (c), there are some abnormal packets due to environmental noise. To mitigate this issue, we adopt an abnormal packet removal approach utilizing the Mahalanobis distance as shown in
\begin{equation}
d(x) = (x - \mu)^T \Sigma^{-1} (x - \mu),
\end{equation}
where $x \in \mathbb{R}^{K}$ is a column vector representing a CSI packet sample with $K$ subcarriers, $ \mu \in \mathbb{R}^{K} $ is the mean vector of the CSI observations, and $ \Sigma \in \mathbb{R}^{ K \times K} $ is the covariance matrix of the subcarrier measurements. The Mahalanobis distance quantifies the deviation of a CSI packet $x$ from a given distribution, incorporating the covariance among vector elements. A CSI packet with a higher Mahalanobis distance indicates an outlier. By eliminating the top 10\% of packets with the largest Mahalanobis distances, we ensure that the remaining CSI packets are suitable for accurate fingerprinting.
\subsubsection{CSI Amplitude Calibration}
\begin{figure}
	\centering
\subfigure{\includegraphics[scale=0.2]{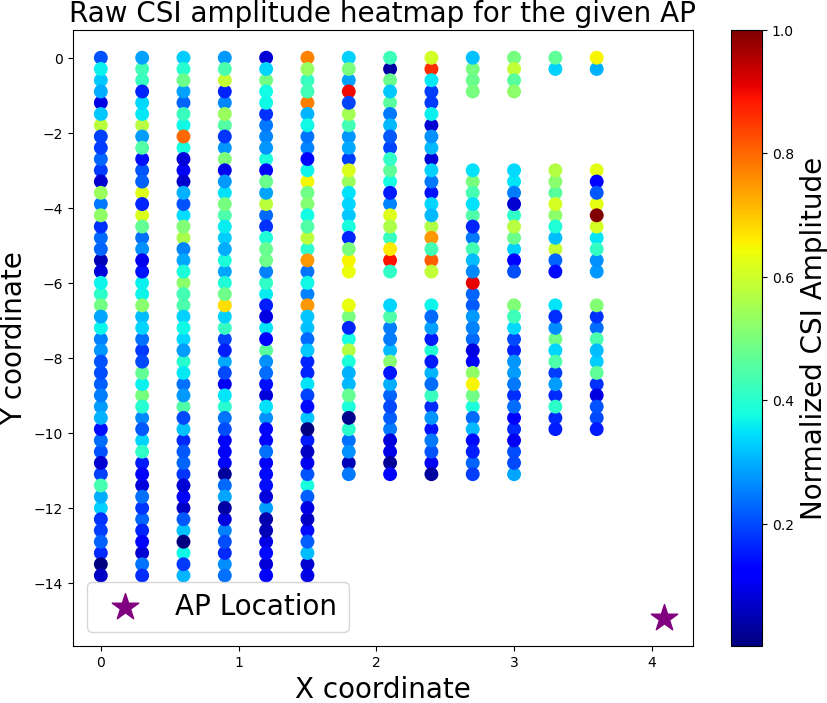}}
\subfigure{\includegraphics[scale=0.2]{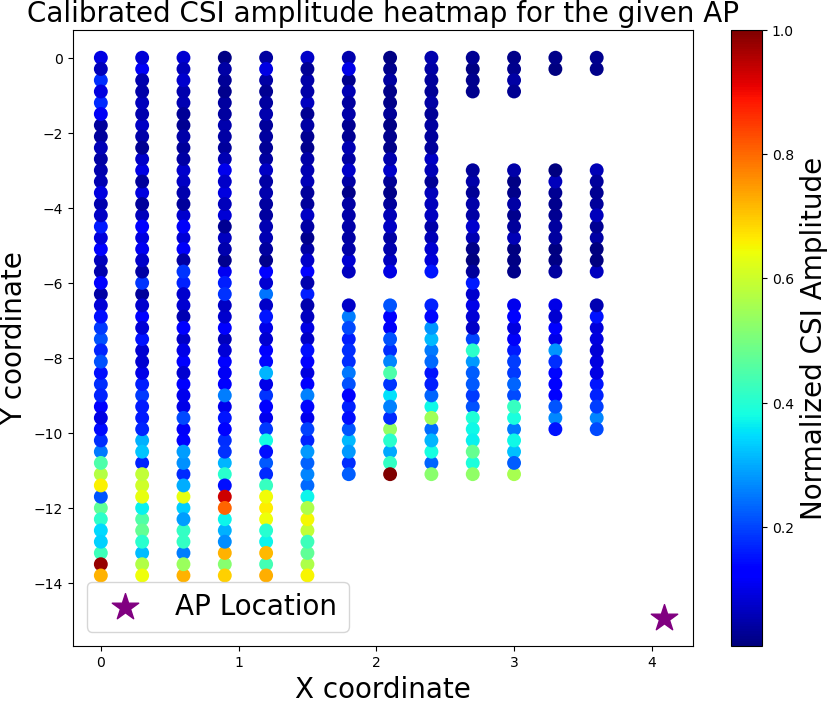}}
	\caption{Left: Raw CSI amplitude heatmap for the given AP; Right: Calibrated CSI amplitude heatmap for the given AP. The given AP marked with a purple star is located at the (4.092,-14.926).}
	\label{fig:amplitude}
\end{figure}
The left subplot of Fig.~\ref{fig:amplitude} shows a heatmap of normalized raw CSI amplitudes transmitted from a mobile device and received by AP1, which is located in the bottom right corner of the area. In contrast to the basic path-loss principle, areas closer to AP1 exhibit lower amplitudes, while some points far from AP1 unexpectedly show higher amplitude values.  This occurs due to the automatic gain control (AGC)~\cite{CRISLoc} at the receiver, which adjusts the CSI amplitude with a factor $s$ that varies according to channel conditions.  To mitigate the impact of AGC on CSI amplitude, we calibrate the raw CSI using RSSI since it is captured before the AGC adjustment and not affected by AGC. Section~\ref{sec:data-format} presents the details of the CSI packet we collected, where each packet header includes an RSSI byte, thus eliminating the need for additional RSSI collection equipment.

As a linear time-invariant system (LTI), AGC uniformly affects all subcarriers. Based on this principle, we apply the adjustment coefficient 
$s$ to all subcarriers of CSI, as proposed in~\cite{CRISLoc}
\begin{equation}
s=\sqrt{\frac{10^{R S SI / 10}}{\sum \left\|H\left(f_k\right)\right\|^2}},
\end{equation}
where $\left\|H\left(f_k\right)\right\|$ represents the amplitude of the 
$k$-th subcarrier extracted from CSI, and RSSI is the received signal strength in dB. By employing this method, raw CSI amplitudes are rescaled to match the corresponding RSSI, effectively mitigating the effects of AGC. The right subplot of Fig.~\ref{fig:amplitude} shows a heatmap of normalized calibrated CSI amplitudes, where the calibrated amplitudes now increase as the distance to AP1 decreases, following the basic path-loss principle. At each location, we select 50\% of the CSI packets after filtering out abnormal ones to calculate the adjustment coefficient $s$, and the remaining packets are adjusted using this coefficient for localization purposes.
\subsubsection{CSI Phase Fitting}
\begin{figure}
    \centering
    \includegraphics[scale=0.45]{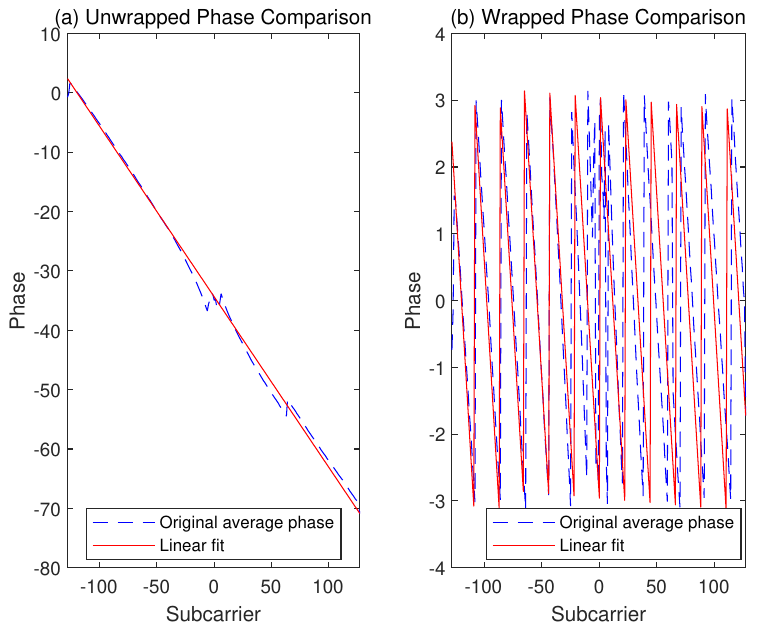}
    \caption{CSI phase processing: (a) unwarpped phase (b) wrapped phase. }
    \label{fig:phase-fitting}
\end{figure}

Figure~\ref{fig:csi-example}
 (b) displays the raw CSI phase for the first packet, spanning the interval $(-\pi,\pi)$. We can see that when the phase exceeds this interval, it results in phase jumps. To address this, we unwrap the raw phase to extend its range from $-\infty$ to $\infty$.  The formula for the unwrapped CSI phase is given by~\cite{8423070}:
\begin{equation}
\phi_k = 2\pi(f_0 + k\Delta f)\frac{d}{c} = (2\pi\Delta f\frac{d}{c})k + 2\pi f_0\frac{d}{c},
\end{equation}
where $d$ represents the path distance, $c$ is the speed of light, $k$ denotes the subcarrier index, and $\Delta f$ is the subcarrier spacing, set at 312.5 KHz for 802.11ac standards.  Utilizing this model, we apply least squares linear regression for estimation. The left subplot in Fig.~\ref{fig:phase-fitting} illustrates the averaged unwrapped phase of all packets, shown with a blue line, while the linear regression fitting result is shown in red. The right subplot compares the original wrapped phases with their linear fits. Following the CSI IFFT transformation, we derive the CIR as a series of signal samples in the time domain as shown in Eq.~\ref{eq:cir}.  According to the IFFT theory~\cite{8423070}, time resolution $\Delta \tau$ of CIR is related to the bandwidth $B$ of the CSI as $\Delta \tau=1 / B$. This relationship indicates that a wider bandwidth improves the resolution of the CIR, allowing for more accurate time-domain channel characterization.

\begin{figure}
    \centering
    \includegraphics[scale=0.35]{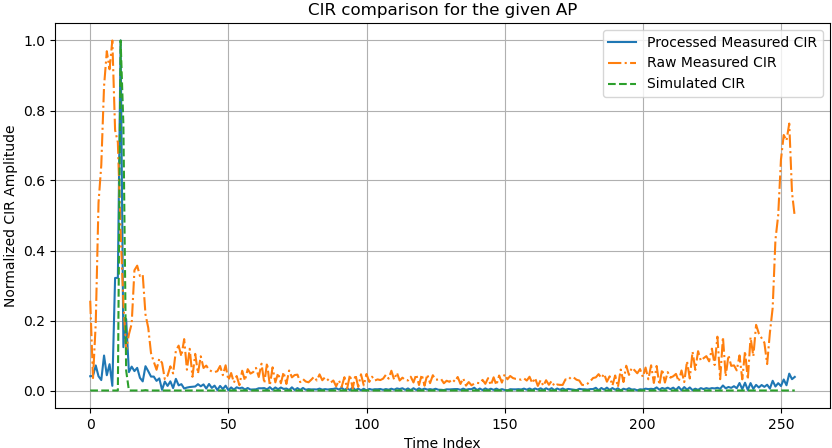}
    \caption{CIR Comparison among transformed by raw CSI IFFT (raw measured CIR marked as orange line), processed CSI IFFT (processed measured CIR marked with blue line), and simulated CIR (marked with green line) generated by the WI platform at the same location.}
    \label{fig:CIR-comparison}
\end{figure}

Figure~\ref{fig:CIR-comparison} presents a comparison of CIRs obtained from both the raw and processed CSI using IFFT. The raw measured CIR is shown by the orange line, while the processed CIR is indicated by the blue line. Additionally, the green line represents the simulated CIR, which serves as the ground truth, generated by the Wireless Insite (WI) platform at the same location for a given AP. The comparison highlights that the CIR derived from unprocessed CSI exhibits multiple noise peaks, especially at the end of the time index. In contrast, the CIR obtained from the processed CSI shows a much cleaner profile, closely matching the simulated CIR, demonstrating the effectiveness of the preprocessing steps.

\subsection{Environmental Data Preprocessing   }~\label{sec:environmantal information}
In this work, environmental data is captured using LiDAR techniques~\cite{luetzenburg2021evaluation}, where the LiDAR sensor emits near-infrared light and measures the time of flight (ToF) of the reflected signal. This process allows for the calculation of distances to objects and then generates the point clouds that reconstruct the 3D environment. The point cloud is formally written as~\cite{10757755}
\begin{equation}
P_{\text{env}} = \{(p_i, s_i) \mid i = 1, \dots, I\},
\end{equation}
where $p_i$ represents the coordinates of the $i$-th point, $s_i$ denotes additional features such as color,  and $I$ is the total number of points in the cloud.

\begin{figure}
    \centering
    \includegraphics[scale=0.75]{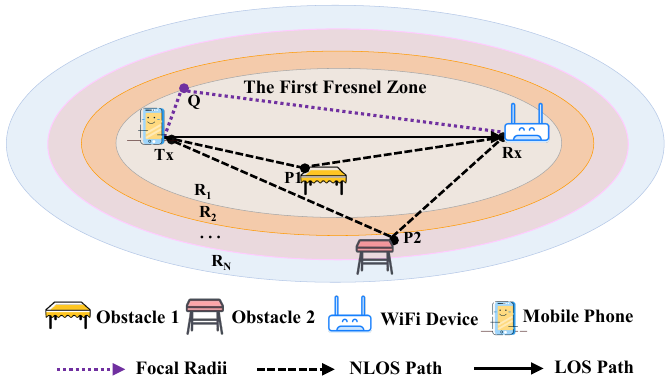}
    \caption{The illustration of the Fresnel zones in an uplink wireless communication system, where $R_n$ represents the $n$-th Fresnel zone with $n=1,2,\dots,N$.}
    \label{fig:tr}
\end{figure}

Meanwhile, we introduce Fresnel zones as the geometric models to interpret the spatial relationships within an environment~\cite{wu2022wifi}. These zones $z_n,n=1,2,\dots,N$ are described as a series of confocal ellipsoids with two foci located at the transmitter (Tx) and receiver (Rx), as illustrated in Fig.~\ref{fig:tr}. Formally, Fresnel zones consisting of $n$ confocal ellipsoids can be defined by the following condition:
\begin{equation}
|T_x Q_n| + |Q_n R_x| - |T_x R_x| = n \cdot \frac{\lambda}{2}, n = 1,\dots,N.
\end{equation}
where $Q_n$ is a point on the $n$-th ellipsoid, $\lambda$ is the wavelength of the electromagnetic wave. The innermost ellipsoid corresponding to $n = 1$ is known as the first Fresnel zone, and the elliptical annuli between successive ellipsoids represent higher-order Fresnel zones. Based on the geometric property of ellipses, where the sum of the distances from any point on an ellipse to its two foci is constant, all points interacting with EM waves on a cross-sectional ellipse of the ellipsoid share the same ToA. Since commercial WiFi devices operate at 5 GHz, corresponding to a wavelength $\lambda$ of approximately 6 cm, the first Fresnel zone forms a relatively narrow ellipsoid around the LOS path. Environmental factors within this zone, such as obstacles, can significantly influence the LOS path by causing diffraction, reflection, and other wave phenomena. These effects in turn can alter signal strength and impact localization accuracy. Consequently, we focus on the environmental information within the first Fresnel zone, as it contains the majority of the signal’s energy due to the LOS path~\cite{yang2024orchloc}.

\begin{figure*}
    \centering 
    \includegraphics[scale=0.32]{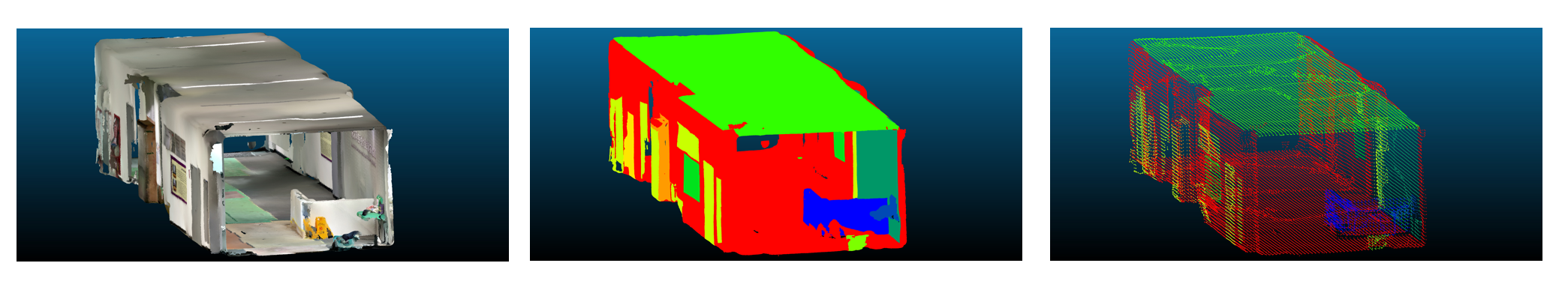}
    \caption{\textbf{Left}: photograph from point cloud; \textbf{Middle}: segmentation with dense point cloud; \textbf{Right}: segmentation with sparse point cloud}
    \label{fig:scenario-point}
\end{figure*}

We use a hall as shown in Fig.~\ref{fig:scenario-point} to demonstrate the extraction of environmental information from a point cloud. The left subfigure displays a photograph reconstructed from the captured point cloud, showcasing the detailed structure of the scene. The middle subfigure illustrates the segmentation of a dense point cloud, where different scene components are color-coded to represent object classes, such as desks in blue and ceilings in green. Although this dense representation offers high precision, it also brings significant computational cost. To address this, the right subfigure shows the segmentation of a sparse point cloud processed via CloudCompare~\cite{girardeau2016cloudcompare},where a minimum point spacing is set, retaining only 0.4\% of the original points. This approach preserves the essential spatial structure and segmentation features while reducing data points and computational load.

\begin{figure*}
    \centering
    \includegraphics[scale=0.8]{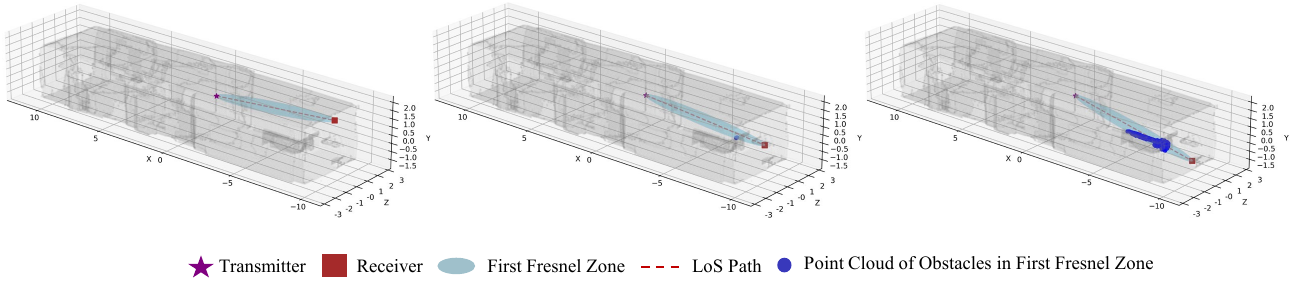}
    \caption{\textbf{Left}: No obstacles in the First Fresnel Zone; \textbf{Middle}: Few obstacles in the First Fresnel Zone; \textbf{Right}: Many obstacles in the First Fresnel Zone, with the LOS path blocked.}

    \label{fig:ffz}
\end{figure*}

Building upon the processed point cloud, Fig.~\ref{fig:ffz} illustrates three cases of the first Fresnel zone in the hall. The transmitter denoted by a star is fixed, while the receiver represented by a square is moved to generate different first fresnel zones. The blue points correspond to the point cloud of the desk. The left subplot shows an unobstructed first Fresnel zone, where the LOS path remains complete, allowing accurate distance estimation during localization. The middle subplot depicts a first fresnel zone with minimal obstacles, where the LOS is still clear but diffraction occurs, leading to interference in the wireless signal and affecting the accuracy of distance estimation. The right subplot illustrates a first Fresnel zone with significant obstacles blocking the LOS path, leading to incorrect distance estimates. These cases highlight the impact of obstacles within the first Fresnel zone on signal propagation and localization accuracy.  To quantify the environmental information, we calculate the proportion of each obstacle within the first Fresnel zone. Obstacles are categorized into $K$ types (e.g., glass, metal, concrete,
plastic, human figures and so on.), and their impact is assessed based on the number of points corresponding to each obstacle. Following this, both the distance and 3D orientation of the transceivers are incorporated, resulting in an environmental information vector \(\mathbf{y} \in \mathbb{R}^{(K+4) \times 1}\).


\section{GenMetaLoc: DESIGN AND IMPLEMENTATION}
In this section, we detail the design and implementation of GenMetaLoc, focusing on three key modules: the meta-learning framework, the environmental information extractor, and the environment-aware diffusion model. Specifically, the meta-learning framework produces fast adaptable model meta-parameters, while the environmental information extractor utilizes point cloud data to capture contextual information. This allows the diffusion model to generate dense CIR fingerprints for new environments with minimal real-world data, effectively eliminating  resource-intensive field data collection.
\subsection{Meta-learning}
Meta-learning is a \emph{learning to learn} approach that enables a model to adapt to new tasks by leveraging experience from related tasks. In this framework, tasks are drawn from a specific distribution, denoted as \(\tau\sim\mathcal{P}(\tau)\), and each task includes a support set for training and a query set for test. In a regression problem considered in this context, a task involves learning a function that maps inputs to continuous outputs based on the samples provided. 

In the meta-training stage, \(M\) training tasks, \(\{\tau_{i}\}_{i=1}^{M} \sim \mathcal{P}(\tau)\), are sampled from the distribution, with corresponding datasets available for the model. Each task consists of input-output pairs in both support and query sets. In the meta-test stage, a new test task \(T \sim \mathcal{P}(\tau)\) is presented, consisting of a small support set and a query set. The objective of meta-learning is to train a model on the \(M\) training tasks, so that it can quickly adapt to the new test task using the small support set and perform well on the query set.

One widely used method in the meta-learning framework is model-agnostic meta-learning (MAML), which aims to learn a good initialization \(\bm{\theta}_{MAML}\) for the model parameters. This initialization enables the model to adapt to new tasks efficiently with only a few gradient updates, leveraging the experience from previous tasks. In the meta-training stage, MAML formulates a meta-optimization problem as: 
\begin{equation}
    \bm{\theta}_{MAML} = \mathop{\arg\min}\limits_{\bm{\theta}} \sum_{i=1}^{M} \mathcal{L}_{\tau_{i}}\left(\bm{\theta} - \alpha \nabla_{\bm{\theta}} \hat{\mathcal{L}}_{\tau_{i}}(\bm{\theta})\right),
\end{equation}
where \(\bm{\theta}\) represents the model parameters, and the loss functions \(\hat{\mathcal{L}}_{\tau_{i}}\) and \(\mathcal{L}_{\tau_{i}}\) are computed based on the support set and query set of the training task \(\tau_{i}\), respectively. The meta-parameters are updated via stochastic gradient descent (SGD) as follows:
\begin{equation}
    \bm{\theta}_{MAML} \gets \bm{\theta}_{MAML} - \beta \nabla_{\bm{\theta}} \sum_{i=1}^{M} \mathcal{L}_{\tau_{i}}\left(\bm{\theta} - \alpha \nabla_{\bm{\theta}} \hat{\mathcal{L}}_{\tau_{i}}(\bm{\theta})\right),
    \label{eq:outer_loop}
\end{equation}
where \(\alpha\) and \(\beta\) denote the step sizes of the inner loop and outer loop, respectively.

\begin{figure*}
    \centering
    \includegraphics[scale=0.7]{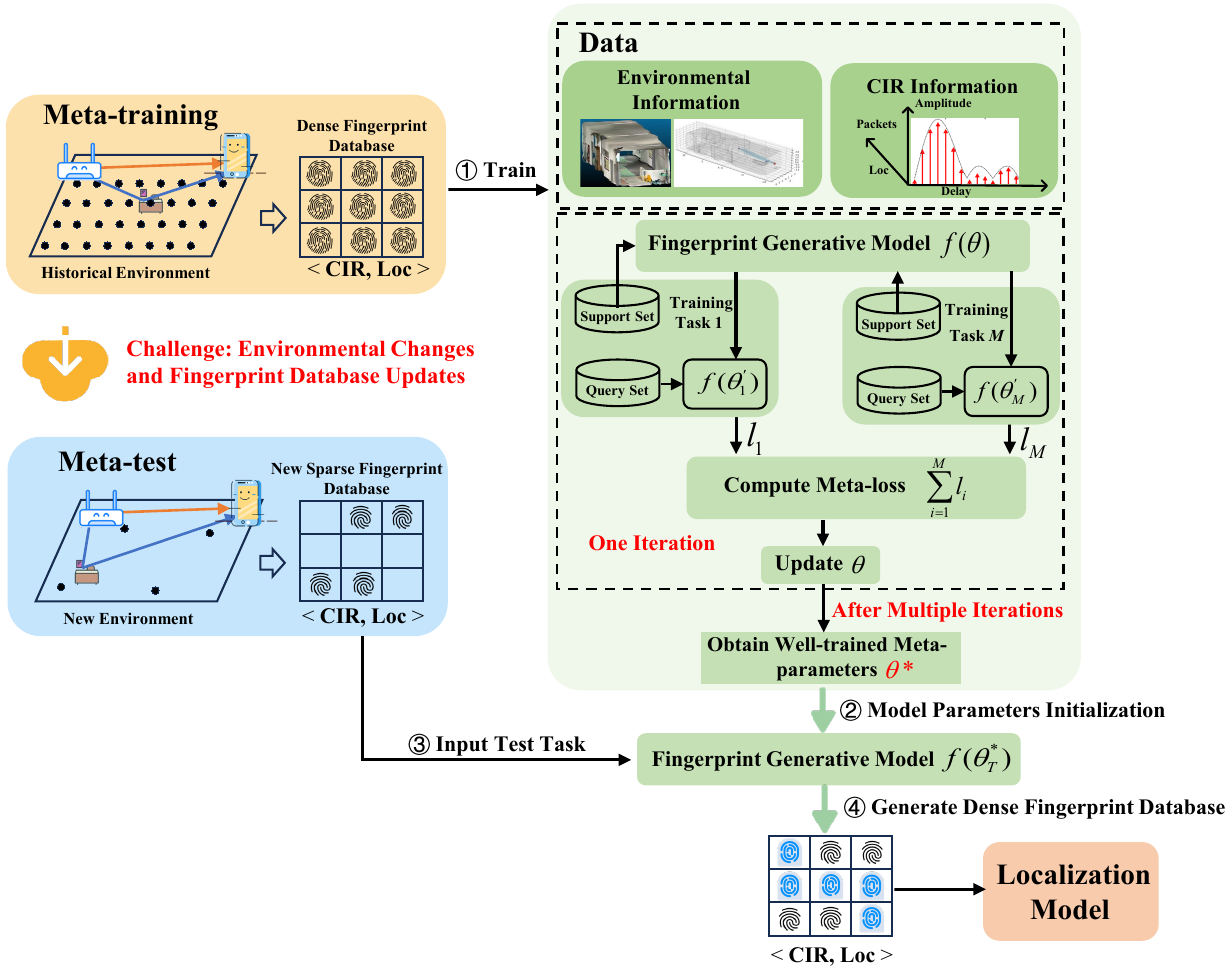}
    \caption{The workflow of the proposed GenMetaLoc framework.}
    \label{fig:workflow}
\end{figure*}

During the meta-test stage, the meta-parameters \(\bm{\theta}_{MAML}\) are fine-tuned to obtain the parameters \(\bm{\theta}_{T}\) for the test task \(T\). This is achieved by updating the meta-parameters using the gradient of the loss function \(\hat{\mathcal{L}}_{T}\left(\bm{\theta}_{MAML}\right)\) computed based on the support set of the test task:
\begin{equation}
    \bm{\theta}_{T} \gets \bm{\theta}_{MAML} - \alpha \nabla_{\bm{\theta}} \hat{\mathcal{L}}_{T}\left(\bm{\theta}_{MAML}\right),
\end{equation}
where the objective is to quickly adapt the model parameters to accurately map from inputs to the corresponding outputs in the query set. Next, we provide a detailed analysis of the MAML framework in our context from perspective of meta-training and meta-test as illustrated in Fig.~\ref{fig:workflow}. 

\subsubsection{Meta-training}
Meta-training is conducted during the offline stage, where dense fingerprints and environmental information from point clouds are collected in the historical environment. The purpose of the meta-training is to derive the well-trained meta-parameters $\bm{\theta}^*$. This is achieved through a generative model that maps the observed environmental information of the points to the corresponding CIR fingerprints. The process is carried out via the following steps: 

Firstly, the generative model is initialized with a predefined network architecture and randomly initialized meta-parameters \(\bm{\theta}\), represented by the parameterized function \(f_{\bm{\theta}}\). A total of \(M\) tasks \(\{\tau_{1}, \dots, \tau_{M}\}\) are sampled from the CIR fingerprint generation task distribution \(p(\tau)\) to serve as training tasks. Each task \(\tau_{i}\) includes a loss function \(\mathcal{L}_{\tau_{i}}\), a support set \(D_{\tau_{i}}^{s}\), and a query set \(D_{\tau_{i}}^{q}\). The loss function \(\mathcal{L}_{\tau_{i}}\) provides task-specific feedback, utilizing mean squared error (MSE) for our regression problem.

Secondly, for each CIR fingerprint generation task \(\tau_{i}\), the model \(f_{\bm{\theta}}\) is trained using the support set \(D_{\tau_{i}}^{s}\), yielding the task-specific loss \(\mathcal{L}_{\tau_{i}}(f_{\bm{\theta}}; D_{\tau_{i}}^{s})\). The task-specific parameters \(\bm{\theta}^{\prime}_i\) are then obtained through a one-step gradient descent update:
\begin{equation}
    \bm{\theta}^{\prime}_i = \bm{\theta} - \alpha \nabla_{\bm{\theta}} \mathcal{L}_{\tau_{i}}(f_{\bm{\theta}}; D_{\tau_{i}}^{s}),
    \label{eq:inner_update}
\end{equation}
where \(\alpha\) is the step size of the inner loop. Since \(\bm{\theta}^{\prime}_{i}\) is derived from a single gradient update on the support set \(D_{\tau_{i}}^{s}\), it provides only limited insight into each task. Further evaluation on the query set \(D_{\tau_{i}}^{q}\) is needed for a more comprehensive understanding of the task performance.

Thirdly, the performance of the one-step update for the \(i\)-th fingerprint generation task is evaluated on the query set using the loss \(\mathcal{L}_{\tau_{i}}(f_{\bm{\theta}^{\prime}_i}; D_{\tau_{i}}^{q})\). The meta-loss is defined as the sum of all task-specific losses, \(\sum_{i=1}^{M} \mathcal{L}_{\tau_{i}}(f_{\bm{\theta}^{\prime}_i}; D_{\tau_{i}}^{q})\), which serves as the meta-objective. We then update the meta-parameters \(\bm{\theta}\) by minimizing the meta-loss:
\begin{equation}
    \bm{\theta}^{*} = \underset{\bm{\theta}}{\arg\min} \sum_{i=1}^{M} \mathcal{L}_{\tau_{i}} \left( f_{\bm{\theta}^{\prime}_i}; D_{\tau_{i}}^{q} \right).
\end{equation}
During meta-optimization, meta-parameters \(\bm{\theta}\) are optimized to achieve the best generalization across all tasks, while  task-specific parameters \(\bm{\theta}_i^{\prime}\) are used to evaluate the performance of the one-step update for each task.

Lastly, the meta-optimization in the outer loop is performed using SGD, updating the meta-parameters \(\bm{\theta}\) as follows:
\begin{equation}
    \bm{\theta} \gets \bm{\theta} - \beta \nabla_{\bm{\theta}} \sum_{i=1}^{M} \mathcal{L}_{\tau_{i}}\left(f_{\bm{\theta}^{\prime}_i}; D_{\tau_{i}}^{q}\right),
    \label{eq:outer_loop}
\end{equation}
where \(\beta\) is the outer loop step size. During this process, the meta-objective is computed using the adapted task-specific parameters \(\bm{\theta}_i^{\prime}\).
\subsubsection{Meta-test}
Meta-test is conducted during the online stage when the environment changes. In this stage, sparse fingerprints are collected in the new environment. After the meta-training, we have obtained well-trained meta-parameters \(\bm{\theta}^*\) that capture the essential characteristics of CIR fingerprints. For an unseen test task \(T \sim \mathcal{P}(\tau)\) in the new environment, with support set \(D_T^{s}\) and query set \(D_T^{q}\), we initialize the generative model \(f_{\bm{\theta}^*}\) with \(\bm{\theta}^*\) and adapt it to the new environment using \(G\) gradient descent steps on \(D_T^{s}\):
\begin{equation}
    \bm{\theta}_{T}^{(n+1)} = \bm{\theta}_{T}^{(n)} - \alpha \nabla_{\bm{\theta}_{T}^{(n)}} \mathcal{L}_T(f_{\bm{\theta}_{T}^{(n)}}; D_T^{s}),
    \label{eq:test_update}
\end{equation}
where \(n = 0, 1, \dots, G-1\), and \(\alpha\) is the step size. After \(G\) updates, the adapted parameters \(\bm{\theta}_T(G)\) are used to evaluate the model on the query set \(D_T^{q}\). The loss function \(\mathcal{L}_T(f_{\bm{\theta}_T(N)}; D_T^{q})\) reflects the model's generalization ability to the new task.

The optimal parameters $\bm{\theta}_{T}^{*}$ for test task $T$ after the meta-test stage are given by 
\begin{equation}
\bm{\theta}_{T}^{*} = \underset{\bm{\theta}_{T}(Q)}{\arg\min} \; \mathcal{L}_{T}\left(f_{\bm{\theta}_{T}(Q)}; D_{T}^{s}\right),
\end{equation}
which results in an adapted generative model for the new environment. This model is then used to generate a dense fingerprint database for the current environment. In this work, we employ a diffusion model as the generative model \(f_{\bm{\theta}}\) for CIR fingerprints, and a detailed discussion of the underlying denoising process is provided in the next section. 
\subsection{Environment-Aware Diffusion Model}~\label{sec:diffusion}
\begin{figure*}
    \centering
    \includegraphics[scale=0.65]{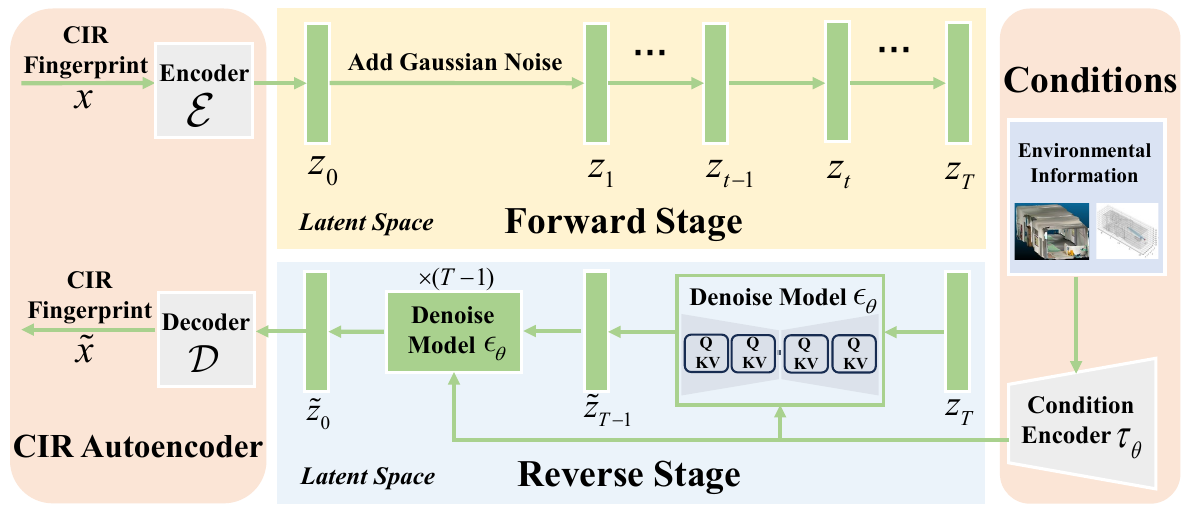}
    \caption{The workflow of the denosing procedure.}
    \label{fig:denoise}
\end{figure*}

Figure \ref{fig:denoise} displays the workflow of the whole environment-aware diffusion model, including the CIR autoencoder, the latent diffusion model with both the forward and reverse stages, and the conditional latent diffusion model. Next, we will provide additional details.
\subsubsection{CIR Autoencoder}
CSI data is inherently high-dimensional, especially when obtained  with the increased bandwidth or massive MIMO setups. For instance, with an 80 MHz bandwidth, up to 256 subcarriers can be obtained, and in massive MIMO systems, the data dimensionality further increases due to the large number of antennas, resulting in high-dimensional CIR data, as discussed in Sec.~\ref{section:process}. To reduce the computational workload, we employ an autoencoder to compress the raw CIR data into a lower-dimensional representation. Specifically, the encoder $\mathcal{E}$ maps the raw CIR fingerprint data $x \in \mathbb{R}^{a \times b \times c}$ to the parameters of a latent distribution, with $a$ representing the number of wireless propagation paths, $b$ representing the number of APs, and $c$ representing the number of antennas:
\begin{equation}
\mathcal{E}(x) = (\mu, \log \sigma^2),
\end{equation}
where $\mu$ and $\log \sigma^2$ denote the mean and logarithmic variance of the latent distribution, respectively. The latent vector $z \in \mathbb{R}^{a^{\prime} \times b^{\prime} \times c^{\prime}}$ is then sampled from this distribution, which is modeled as:
\begin{equation}
z \sim \mathcal{N}(\mu,\sigma^2).\label{eq:latent}
\end{equation}
During the encoding process, a reparameterization trick is applied to obtain the latent vector $z$:
\begin{equation}
z = \mu + \epsilon \cdot \sigma, \quad \epsilon \sim \mathcal{N}(0, \mathbf{I}),
\end{equation}
where $\epsilon$ is sampled from a standard normal distribution. The purpose of this trick is to make the sampling process differentiable, which allows gradients to be backpropagated through the model parameters ($\mu$ and $\sigma$), enabling the efficient model training.
The decoder $\mathcal{D}$ then reconstructs the raw CIR from the latent vector $z$, which yields
\begin{equation}
\tilde{x} = \mathcal{D}(z).
\end{equation}
The objective of the autoencoder is to minimize the reconstruction loss, defined as the sum of the mean squared error (MSE) and the Kullback-Leibler divergence (KLD):
\begin{align}
\mathcal{L}_{\text{recon}} &= \text{MSE}(\tilde{x}, x) \notag \\
&+ \text{KLD}(\mathcal{N}(\mu, \sigma^2) \parallel \mathcal{N}(0, \mathbf{I}))
\end{align}

where the MSE aims to ensure that the reconstructed CIR data closely matches the original input data, thereby preserving the critical features of the CIR. Besides, in order to avoid arbitrarily high-variance latent spaces, we imposes a slight KL-penalty term towards a standard normal on the learned latent. By this procedure, we obtain a compressed latent variable $z$ that replaces the original CIR fingerprint $x$ in the subsequent diffusion steps, thereby saving computational resources.
\subsubsection{Latent Diffusion Model}
Diffusion models are probabilistic models designed to learn the distribution of data \( p(x) \), by progressively denoising a normally distributed variable. In the case of latent diffusion model, we consider a latent data distribution \( p(z) \), where \( z \) represents the compressed latent variable of original CIR data $x$ as discussed in Eq.~\ref{eq:latent}. The learning process can be divided into two main stages: the forward (noise addition) stage and the reverse (denoising) stage.

In the forward stage, we gradually add Gaussian noise to the latent variable of the original CIR data $z_0$ over $T$ time steps, resulting in increasingly noisy versions of the data. This process is modeled as a fixed Markov chain and transforms $z_0$ into a noisy version $z_T$ that follows a standard Gaussian distribution. The forward process can be expressed as
\begin{equation}
z_t = z_{t-1} + \epsilon_t, \quad \epsilon_t \sim \mathcal{N}(0, \sigma_t^2), \quad t = 1, \dots, T,
\end{equation}
where at each step $t$, a Gaussian noise $\epsilon_t$ is added to obtain a noisy version of the latent variable.

In the reverse stage, the denoising model $f_\theta = \epsilon_{\theta}(z_t,t)$ progressively denoises the data from $z_T$ to $z_0$ by predicting and removing noise at each step $t$ with $t = T, \dots, 1$. The predicted  latent data at each step $t$ is denoted by $\tilde{z}_t$, which allows $\tilde{z}_{t-1}$ to be estimated as: 
\begin{equation}
\tilde{z}_{t-1} = \tilde{z}_t -\epsilon_{\theta}(z_t,t),
\end{equation}
and this process continues iteratively until $\tilde{z}_0$ is obtained.

The training objective for the diffusion model is to minimize the difference between the predicted noise and the actual noise, expressed as:
\begin{equation}
L_{LDM} = \mathbb{E}_{z_0, \epsilon \sim \mathcal{N}(0, 1), t}\left[\left\|\epsilon - \epsilon_\theta(z_t, t)\right\|_2^2\right], ~\label{eq:dm}  
\end{equation}
where the goal is to ensure that the model accurately predicts the noise at each step, thereby enabling it to effectively reverse the diffusion process and recover the original latent data $z_0$.
\subsubsection{Conditional Latent Diffusion Model}
To better understand conditioning in diffusion models, consider ChatGPT generating a response based on a prompt like "Explain the concept of diffusion models." These prompts serve as conditions that guide the generative process, ensuring the generated output aligns with the description.  Specifically, latent diffusion models can formalize conditional distributions as $p(z\mid y)$, where $y$ represents the conditioning input. In our context, the conditioning information $y$ represents environmental data derived from point clouds as we described in Sec.~\ref{sec:environmantal information}.

To make diffusion models more adaptable for conditional generation, we augment their architecture by incorporating a cross-attention mechanism into the U-Net backbone. Specifically, to process conditioning inputs \(y\) from various modalities, we introduce an environment-specific encoder \(\tau_\theta\), which projects \(y\) into an intermediate representation \(\tau_\theta(y) \in \mathbb{R}^{M \times d_\tau}\), where \(M\) represents the number of elements in the input \(y\) after encoding, and \(d_\tau\) is the dimension of each element in the intermediate representation.  This cross-attention allows the model to dynamically integrate information from the conditioning inputs, guiding the denoising process at each stage and ensuring the output aligns with the given conditions.  This representation is then mapped into the intermediate layers of the U-Net through a cross-attention layer implementing 
\begin{equation}
\operatorname{Attention}(Q, K, V)=\operatorname{softmax}\left(\frac{Q K^T}{\sqrt{d}}\right) \cdot V,
\end{equation}
with
\[
Q=W_Q^{(i)} \cdot \varphi_i\left(z_t\right), \quad K=W_K^{(i)} \cdot \tau_\theta(y), \quad V=W_V^{(i)} \cdot \tau_\theta(y),
\]
where \(\varphi_i\left(z_t\right) \in \mathbb{R}^{F \times d_\epsilon^i}\) represents a flattened intermediate representation of the U-Net, with \(F\) representing the number of spatial elements  in the U-Net feature map at layer \(i\), and \(d_\epsilon^i\) is the dimension of the feature vector for each spatial element. \(W_Q^{(i)}, W_K^{(i)}, W_V^{(i)}\) are learnable projection matrices. The scaling factor \(\sqrt{d}\) maintains stability during the training process, where \(d\) is the feature size, which ensures appropriate scaling of the dot-product values during attention computation.

Based on this, Eq.~(\ref{eq:dm}) can be rewritten as  
\begin{equation}
L_{CLDM}=\mathbb{E}_{\mathcal{E}(x), y, \epsilon \sim \mathcal{N}(0,1), t}\left[\left\|\epsilon-\epsilon_\theta\left(z_t, t, \tau_\theta(y)\right)\right\|_2^2\right],
\end{equation}
where both  $\tau_\theta$ and $\epsilon_\theta$ are optimized jointly. 
By leveraging cross-attention and a domain-specific encoder, we transform the diffusion model into a data source for generating CIR conditioned on diverse environmental inputs. This approach facilitates the sim-to-real transition by training the localization model with synthetic CIR data from the diffusion model, which is then applied to real-world environments, reducing the need for extensive field data collection.

\section{Experimental Setup}
This section presents the data collection platform, data format, experimental scenarios, and baseline models used for comparison. The datasets are collected from several real-world environments through site surveys conducted on the campus of The Chinese University of Hong Kong, Shenzhen.
\begin{figure}
    \centering
    \includegraphics[scale=0.45]{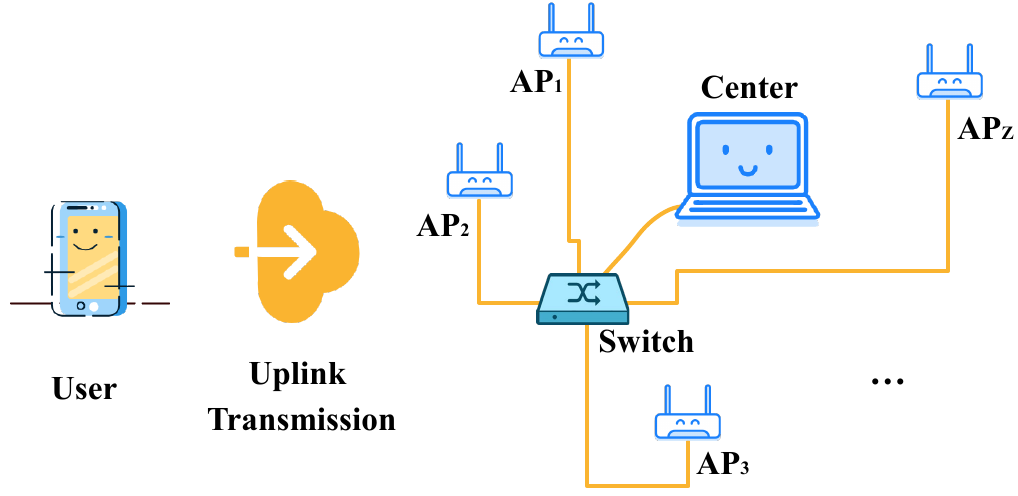}
    \caption{The overall uplink communication platform consisting of $Z$  APs as receivers and a mobile phone as the transmitter.}
    \label{fig:uplink-communication-framework}
\end{figure}
\begin{figure*}
  \centering
  \subfigure[]{%caption of the subfloat
  \includegraphics[scale=0.06]{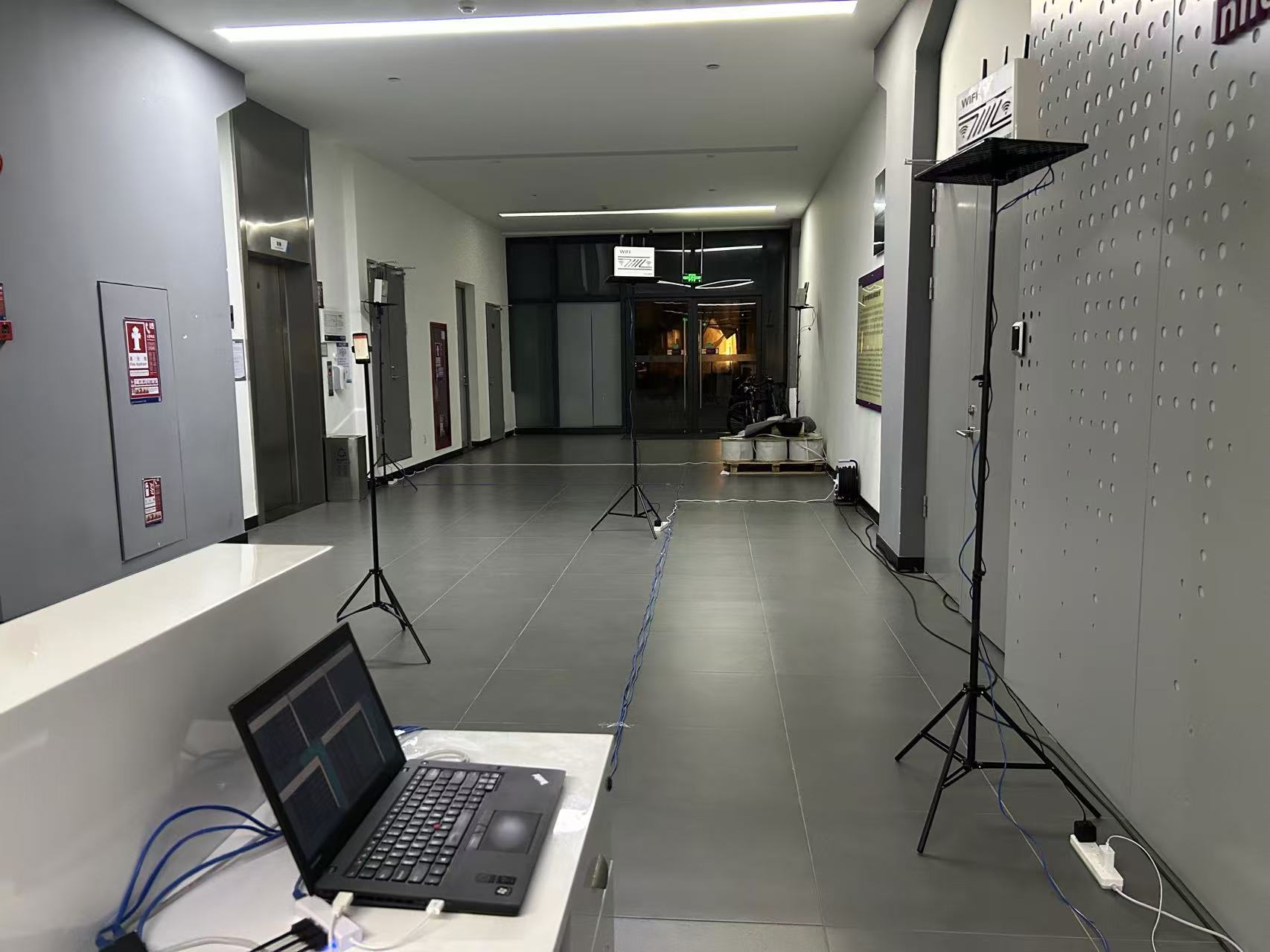}}
  \subfigure[]{%caption of the subfloat
  \includegraphics[scale=0.21]{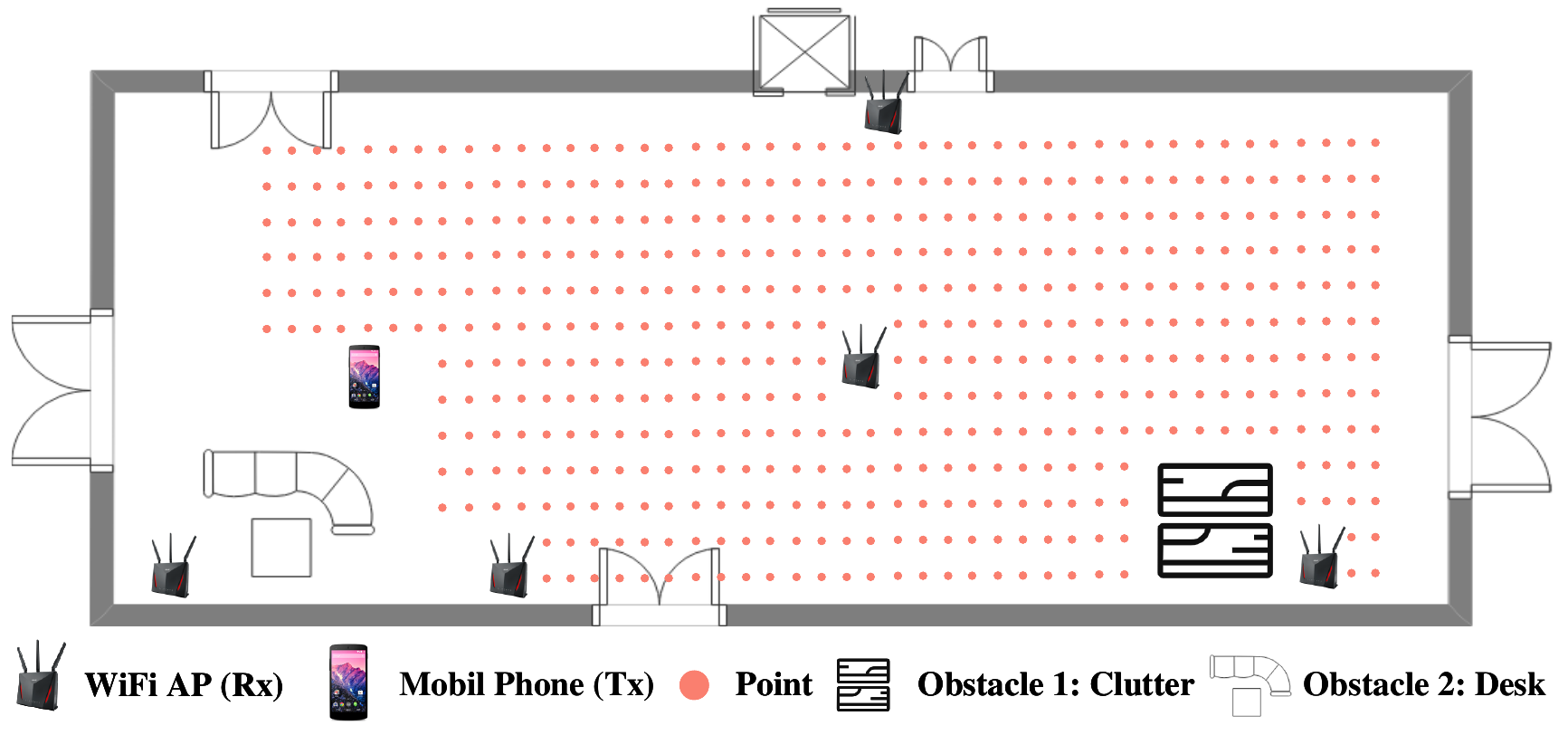}}
  \subfigure[]{%caption of the subfloat
  \includegraphics[scale=0.025]{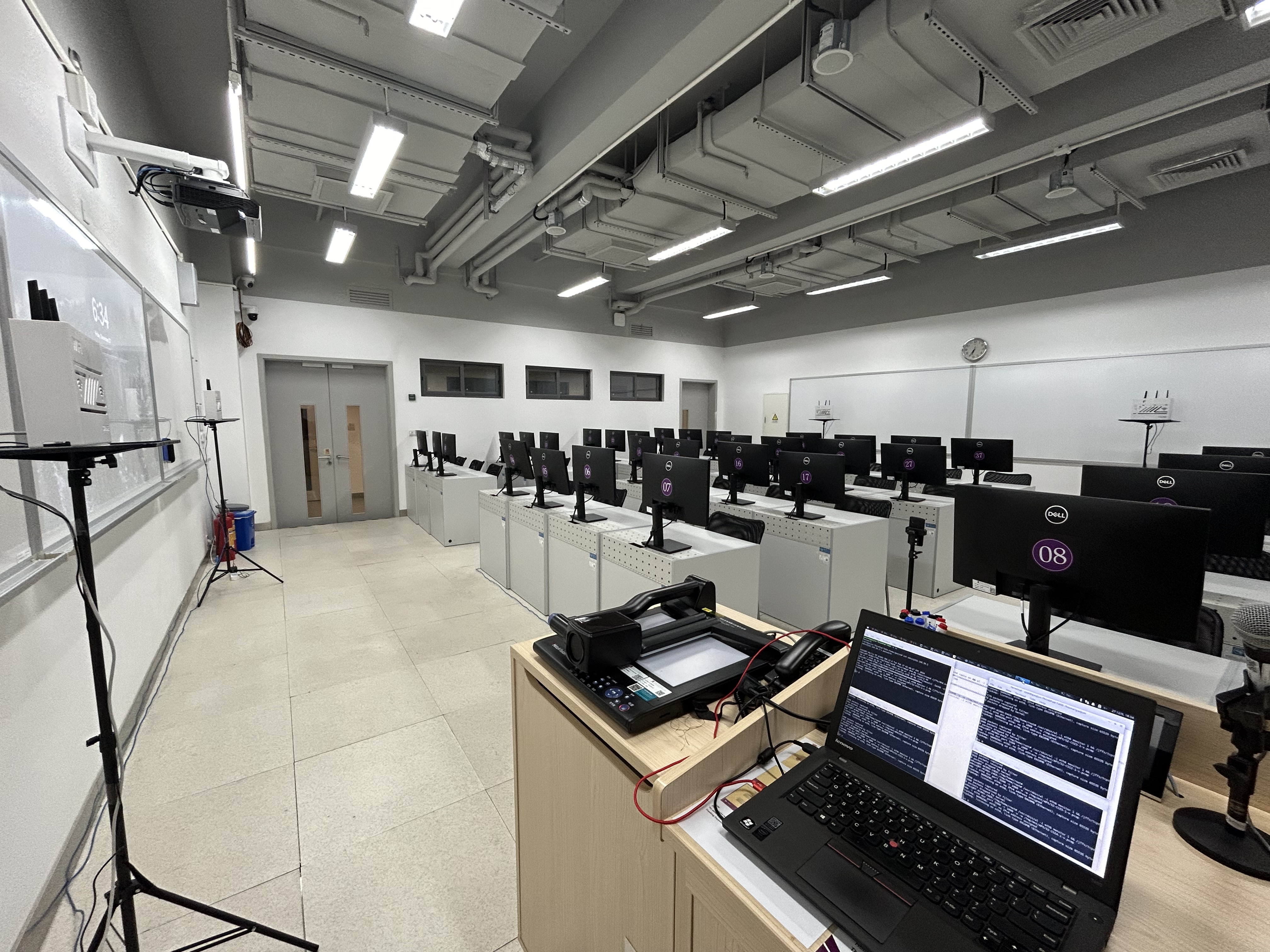}}
  \subfigure[]{%caption of the subfloat
  \includegraphics[scale=0.1]{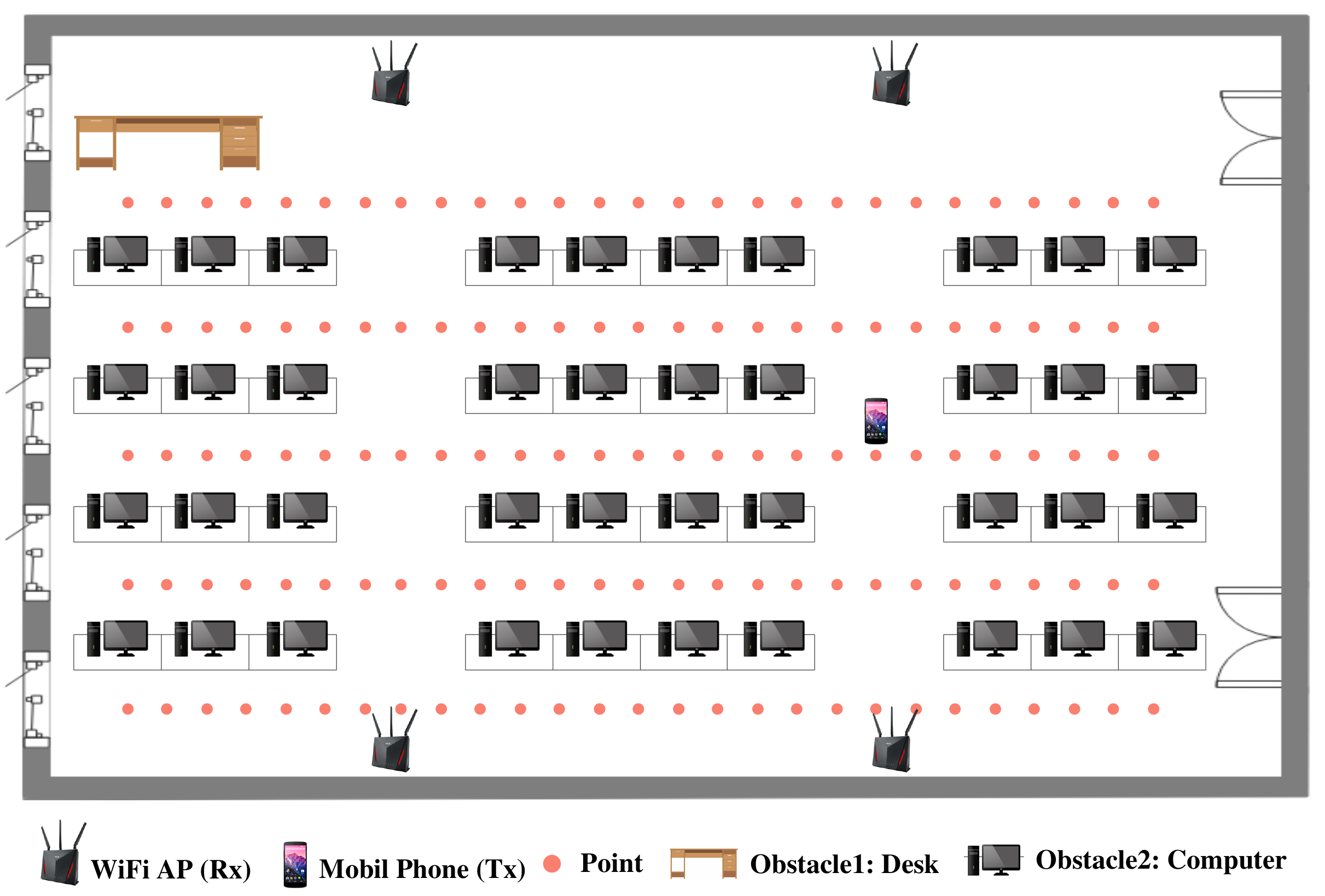}
  }
\caption{Photographs and layouts of the hall and the lab. (a) The photograph of the hall. (b) The layout of the hall. (c) The photograph of the lab. (d) The layout of the lab. }
\label{fig:scenarios}
\end{figure*}
\subsection{Data Collection Platform}
As shown in Fig.~\ref{fig:uplink-communication-framework}, we built an uplink communication platform consisting of $Z=5$ APs connected to a central server via a network switch. Each AP is assigned a unique IP address, enabling independent and simultaneous data reception from the mobile phone. Data collection is performed using the Nexmon CSI Extractor Tool~\cite{nexmon:project}, an open-source project that enables per-frame CSI extraction. The tool supports up to four spatial streams and four receiving chains on Broadcom and Cypress WiFi chips, with bandwidths up to 80 MHz in both the 2.4 GHz and 5 GHz bands. It is compatible with a variety of devices, including low-cost platforms like the Raspberry Pi, mobile devices such as Nexus smartphones, and advanced WiFi APs~\cite{gringoli2019free}. 

In our experimental platform, a Nexus 5 smartphone is used as the transmitter, while ASUS RT-AC86U routers act as the receivers. The Nexmon framework\cite{nexmon:project} is applied to patch a modified firmware\cite{gringoli2019free}, enabling the extraction of RSSI and CSI data. The modified firmware embeds the collected data into a User Datagram Protocol (UDP) stream, which is captured in .pcap format using the \textit{tcpdump}
 utility. Since the WiFi functionality is disabled during the data collection process, an Ethernet link is employed to ensure communication between each WiFi AP and the mobile phone.

The system operates under the 802.11ac standard, in the 5 GHz band on channel 157 (5785 MHz) with an 80 MHz bandwidth. In the S1 LOS-dominant hall, the mobile phone is placed at a height of 1.5 meters, and each AP is mounted on a 2-meter-high stand to ensure the signal coverage. In the S2 lab, to create a NLOS-dominant environment, the mobile phone is positioned at 1.2 meters, and each AP is placed on a 1.5-meter-high stand. At each location, 900 CSI packets are collected to ensure sufficient data for rigorous analysis and robust evaluation.
\subsection{Data Format}~\label{sec:data-format}
\begin{figure}[H]  % This tells LaTeX to place the figure exactly here
    \centering
    \includegraphics[scale=0.35]{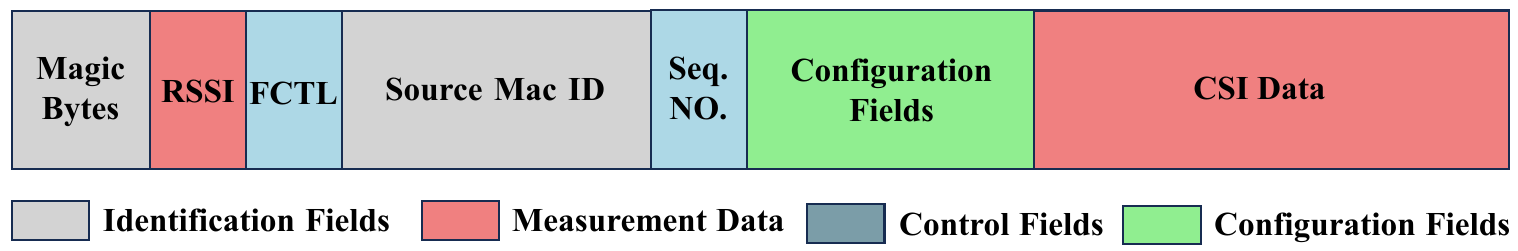}
    \caption{A portion of the collected data packet format with RSSI and CSI fields highlighted in red color.}
    \label{fig:frame}
\end{figure}

 Figure~\ref{fig:frame} presents a portion of the collected data packet\footnote{The complete data packet includes the packet header (12 bytes), network headers (Ethernet, IP, and UDP, totaling 46 bytes), Nexmon Metadata (18 bytes), and CSI Data ($n_{\text{sub}} \times 4 \, \text{bytes}$, where \( n_{\text{sub}} \) represents the number of subcarriers, and each subcarrier contains 4 bytes~\cite{voggu2021decimeter,nexcsi}). Here, we focus on the portion containing the required RSSI and CSI data.}, which is structured into four categories: identification fields, control fields, configuration fields, and measurement data. The identification fields, such as Magic Bytes and Source Mac ID, ensure packet recognition and source verification. The control fields manage packet flow and maintain transmission order. The configuration fields, comprising Core and Spatial Stream, ChanSpec, and Chip Version, provide critical information on channel specifications, hardware capabilities, and spatial stream configurations for accurate data interpretation. Finally, the measurement data, including the RSSI and CSI that we focus on, have been detailed in the previous section.

Based on the packet structure, we extract the required RSSI and CSI data, which serve as the foundation for subsequent processing and analysis as introduced in Sec.~\ref{section:process}. Together, these fields enable efficient packet management, device identification, and accurate decoding of wireless communication parameters, ensuring reliable data transmission.
\subsection{Experimental Scenarios}
\begin{figure}[H]
\centering
\subfigure[]{\includegraphics[scale=0.201]{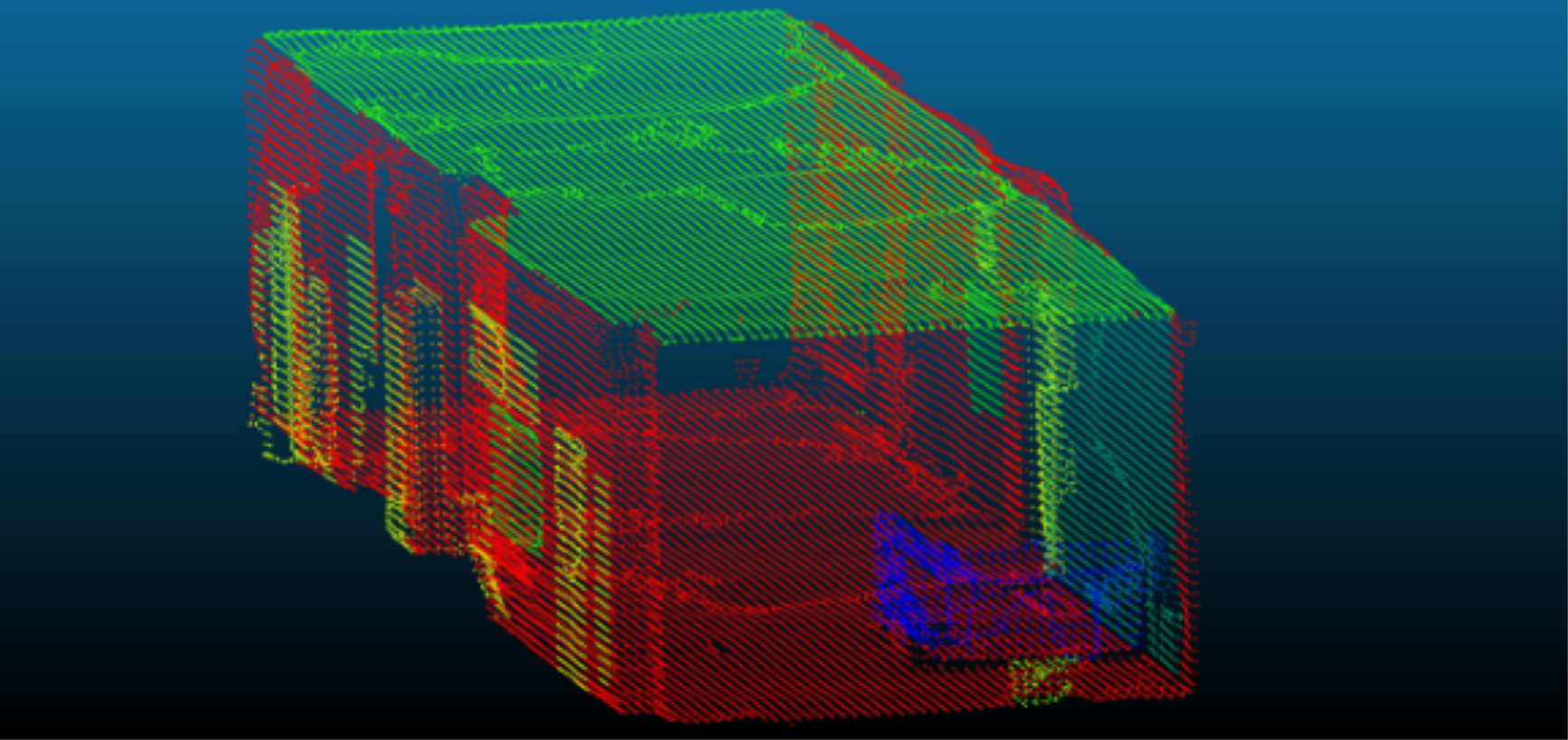}}
\subfigure[]{\includegraphics[scale=0.236]{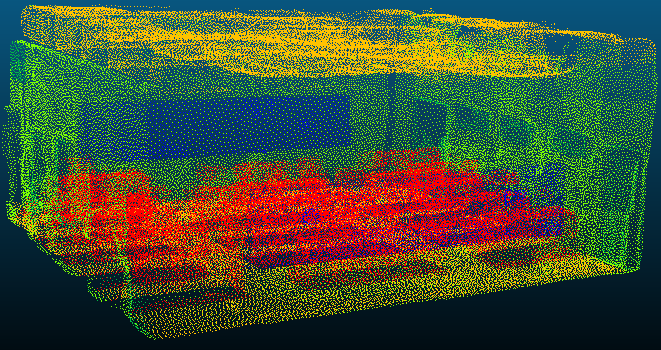}}
\caption{(a) Point cloud representation in the hall. (b) Point cloud representation in the lab. We perform a rough segmentation of obstacles based on common material types using different colors.}~\label{fig:point-cloud}
\end{figure}

As illustrated in Fig.~\ref{fig:scenarios}, the experiments are conducted in two different scenarios:  the hall and the lab. Specifically, as shown in Fig. ~\ref{fig:scenarios} (a) and (b), the hall is a 20 m $\times$ 5 m area with a largely open layout, making it a LOS-dominant environment. However, human activity, such as people walking or elevators opening and closing, introduces minor variations. Data was collected from 500 points spaced 0.3 m apart at five AP locations. On the other hand, as shown in Fig. ~\ref{fig:scenarios} (c) and (d), the lab is a 9 m $\times$ 8 m area with numerous obstacles, such as desks and computers, creating a NLOS-dominant environment. Data was collected from 135 points spaced 0.3 m apart at four AP locations. Point clouds representing environmental information were captured from these scenarios using a 3D scanning application on an iPad Pro device as shown in Fig.~\ref{fig:point-cloud}. Due to space limitations, we only display a particular point cloud image for each scenario. We perform a rough segmentation of obstacles based on common material types, using different colors to represent various materials such as glass, metal, concrete, plastic, and human figures.
\subsection{Baseline Models}
The environment-aware diffusion model consists of two downsampling layers, a cross-attention layer at the bottleneck for integrating environmental data, followed by two upsampling layers, and a final convolution layer to refine the outputs. During the meta-training stage, the step size of the inner loop $\alpha$ and the outer loop $\beta$ are set to 0.01 and 0.001, respectively. In addition, we set the number of gradient descent steps for the inner loop to 5. We compare our proposed framework GenMetaLoc against state-of-the-art methods as detailed below:
\begin{itemize}
\item GenMetaLoc using CSI amplitude only: To verify the effectiveness of the proposed data processing methods, we remain the neural network architecture unchanged from the proposed framework, utilizing only the CSI amplitude for training.
\item {Random Initialization (RI)}: We randomly generate a set of network parameters for initializing the new environment. The task format and the hyperparameters of the neural network remain the same as those used in the GenMetaLoc.
\item KNN~\cite{bahl2000radar}: We adopt the Euclidean distance metric to select the closest $K=5$ RPs for the estimated TP in the signal space. The averaged locations of the selected RPs are then treated as the estimation result.
\item {MetaLoc~\cite{10274764}}: The localization model employs a deep neural network that takes wireless fingerprints as input and outputs corresponding location coordinates.  Meta-parameters are trained with historical data to initialize the network. In contrast to GenMetaLoc, this model does not incorporate environmental information.
\item {ConFi~\cite{chen2017confi}}: We keep the neural network architecture and dataset the same as the proposed method. Data from historical environments serve as the training set, while data collected in new environment serve as the test set.
\end{itemize}

\section{Experimental Results}
In this section, we formulate the environment-aware fingerprint generation as a regression problem and present results based on real-world data collected from a LOS-dominant hall and a NLOS-dominant lab as we introduced in Fig.~\ref{fig:scenarios}.
\begin{figure}[H]
    \centering
    \includegraphics[scale=0.41]{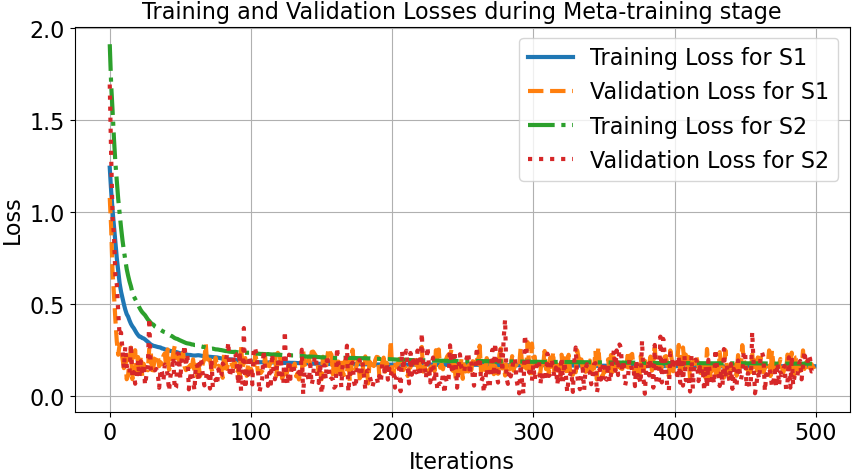}
    \caption{The convergence comparison between training loss and validation loss during the meta-training stage. Specifically, the blue and green curves represent the training loss for S1 and S2, respectively, while the orange and red curves show the validation loss for S1 and S2 during the meta-training stage. All curves converge, indicating effective model training without overfitting.}
    \label{fig:meta-train-loss}
\end{figure}
\subsection{Convergence Analysis}
Figure~\ref{fig:meta-train-loss} shows the convergence comparison between training loss and validation loss during the meta-training stage for S1 and S2 scenarios. All loss curves converge, indicating that the meta-parameters of the CIR fingerprint-based generative model are effectively optimized in the historical environments. The training loss for S1 converges faster than for S2, and the validation loss for S2 exhibits higher variance. This difference is due to S2 being a more complex NLOS-dominant environment, which results in noisier wireless data compared to the LOS-dominant S1 scenario. 

\begin{figure*}
    \centering
    \subfigure[]{\includegraphics[scale=0.3]{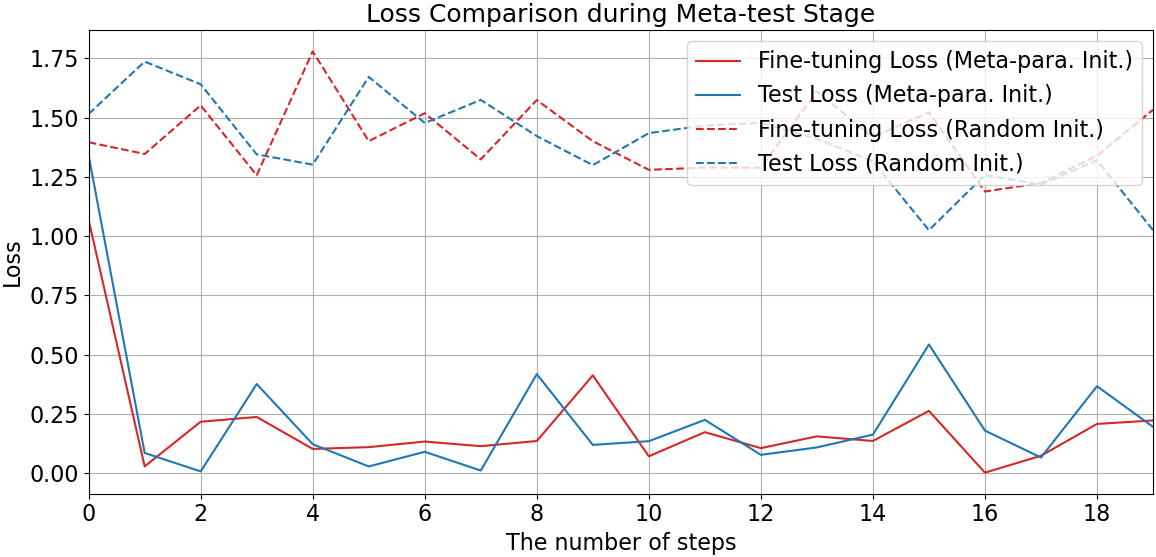}}
    \subfigure[]{\includegraphics[scale=0.3]
    {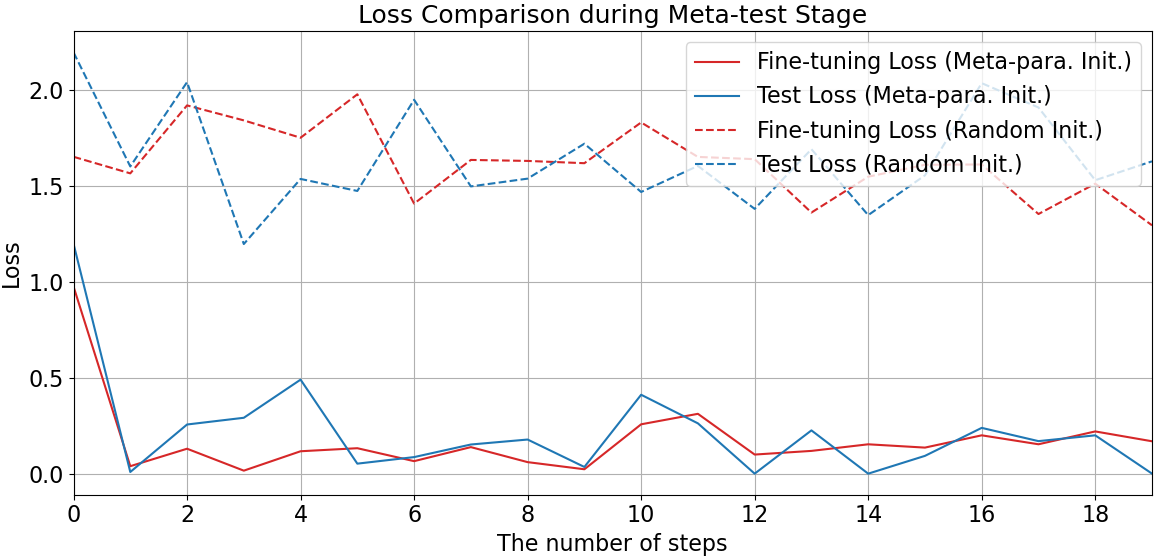}}
    \caption{(a) Loss comparison during the meta-test stage in the S1 scenario. (b) Loss comparison during the meta-test stage in the S2 scenario. Specifically, the solid red and blue curves represent fine-tuning and test loss with meta-parameters initialization, while the dashed red and blue curves represent the corresponding losses with random initialization. Notably, meta-parameters initialization enables faster adaptation to new environments, achieving significantly lower loss than random initialization in both S1 and S2.}
    \label{fig:meta-test-loss}
\end{figure*}

When the environments change, Fig.~\ref{fig:meta-test-loss} (a) and (b) demonstrate the convergence during the meta-test stage in the S1 and S2 scenarios, respectively. The solid red and blue curves represent the fine-tuning loss and test loss for the fingerprint-based generative model initialized with well-trained meta-parameters, while the dashed curves correspond to random initialization. The results show that meta-parameters initialization enables the CIR generative model to adapt faster to new environments and achieve significantly lower loss compared to random initialization. In our previous work~\cite{gao2022metaloc}, we have verified that the random initialization method depends heavily on the amount of training data, with performance improving as the data scale increases. When the data size reaches roughly 60 times the amount required by GenMetaLoc, its performance stabilizes and becomes comparable to that of GenMetaLoc, which only needs 10 samples for training without overfitting.  Furthermore, according to the Theorem 1 in~\cite{zhou2021task}, the rapid adaptation of GenMetaLoc is attributed to the smaller Euclidean distance between the well-trained meta-parameters $\bm{\theta}^{*}$ and the optimal parameters $\bm{\theta}_T^{*}$ for the test task $T$, which leads to a smaller excess risk and ensures good test performance.
\begin{figure*}
	\centering
\subfigure[]{\includegraphics[scale=0.388 ]{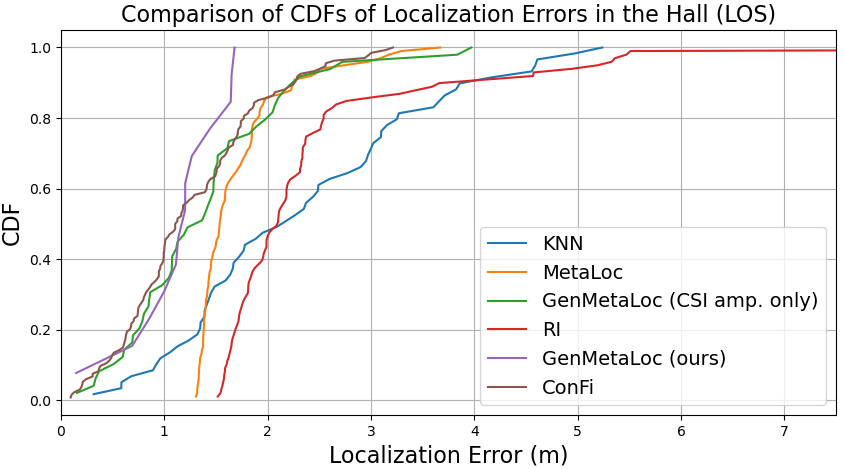}}
\subfigure[]{\includegraphics[scale=0.384]{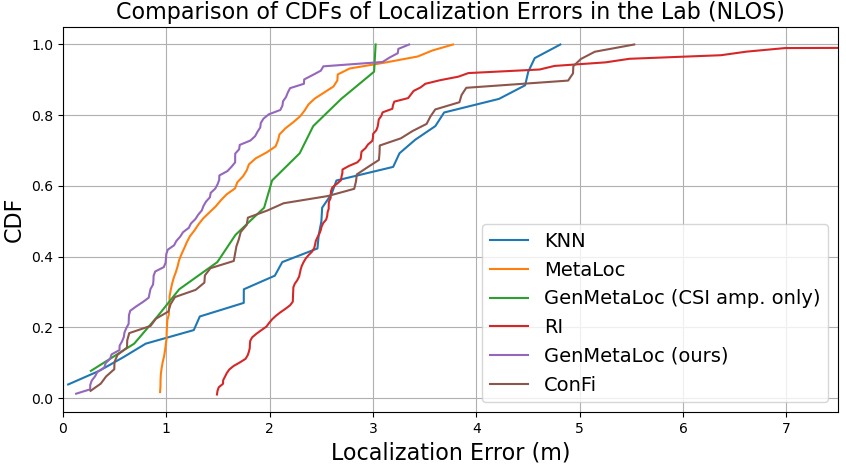}}
\caption{(a) Comparison of CDFs of localization errors in the hall (LOS) across different methods. (b) Comparison of CDFs of localization errors in the lab (NLOS) across different methods.}~\label{fig:cdf}
\end{figure*}
\begin{table*} 
\caption{Localization results in the new environment using real data collected from the site surveys}
\centering
\begin{tabular}{ccccc} 
\toprule[1.5pt] 
\multicolumn{1}{c}{\multirow{2}*{\textbf{Methods}}}& \multicolumn{2}{c}{\textbf{Hall (LOS)}}&\multicolumn{2}{c}{\textbf{Lab (NLOS)}}\\
\multicolumn{1}{c}{}&\textbf{Mean Error (m)}&\textbf{80th Percentile Errors (m)
}&\textbf{Mean Error (m)}&\textbf{80th Percentile Errors (m)}\\  
\hline 
\multicolumn{1}{c}{\textbf{GenMetaLoc (ours)}}                     & \multicolumn{1}{c}{\textbf{1.21}}          & \multicolumn{1}{c}{\textbf{1.65}}        & \multicolumn{1}{c}{\textbf{1.36}}         & \multicolumn{1}{c}{\textbf{1.98}}           \\
\multicolumn{1}{c}{GenMetaLoc (CSI amp. only)}                     & \multicolumn{1}{c}{1.43}          & \multicolumn{1}{c}{2.05}        & \multicolumn{1}{c}{1.69}        & \multicolumn{1}{c}{2.53}   \\       
\multicolumn{1}{c}{MetaLoc}                     & \multicolumn{1}{c}{1.81}          & \multicolumn{1}{c}{1.92}         & \multicolumn{1}{c}{1.77}         & \multicolumn{1}{c}{2.35}          \\
\multicolumn{1}{c}{RI}                     & \multicolumn{1}{c}{2.51}          & \multicolumn{1}{c}{2.55}  & \multicolumn{1}{c}{2.89}&\multicolumn{1}{c}{3.11}         \\
\multicolumn{1}{c}{ConFi}                     & \multicolumn{1}{c}{1.27}          & \multicolumn{1}{c}{1.78} & \multicolumn{1}{c}{2.42}& \multicolumn{1}{c}{3.52}        \\
\multicolumn{1}{c}{KNN}                     & \multicolumn{1}{c}{2.37}          & \multicolumn{1}{c}{3.34}  & \multicolumn{1}{c}{2.59}&\multicolumn{1}{c}{3.67} \\
\bottomrule[1.5pt]
\end{tabular}
\label{tab:accuracy}
\end{table*}
\subsection{Localization Accuracy}
\begin{figure*}
	\centering
\subfigure[]{\includegraphics[scale=0.385 ]{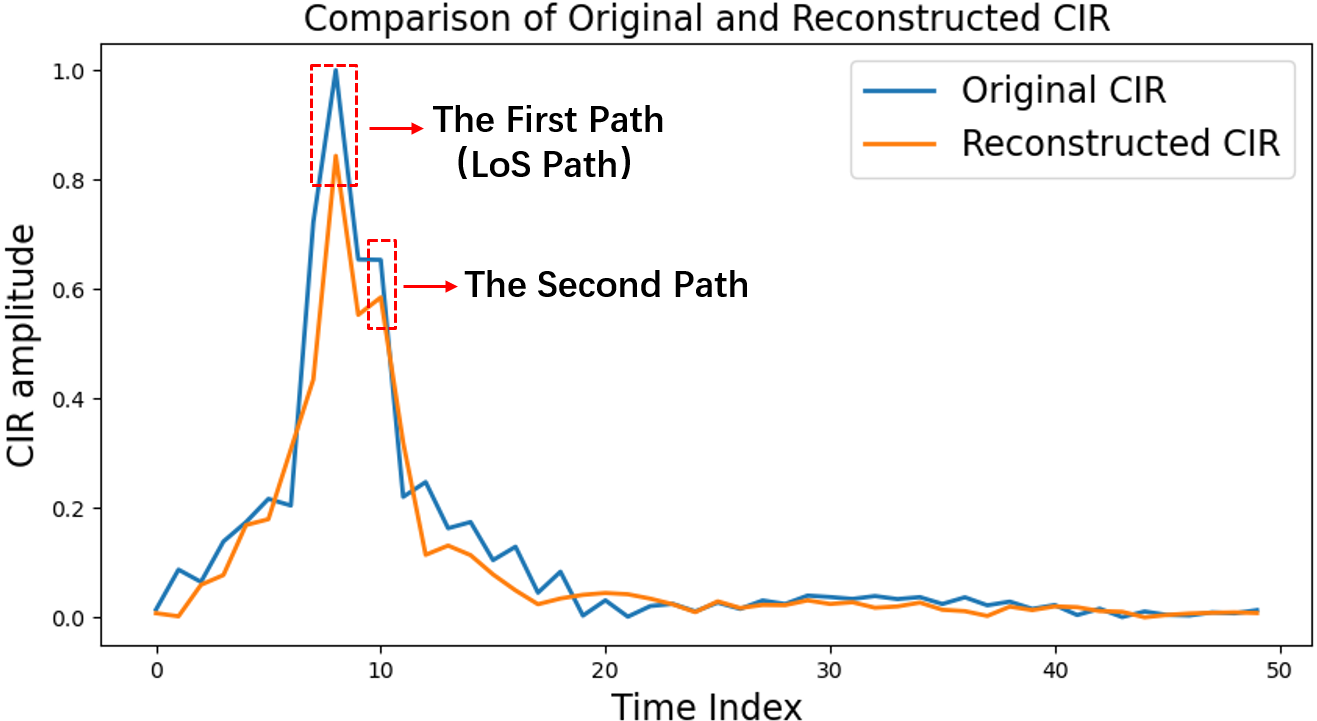}}
\subfigure[]{\includegraphics[scale=0.385]{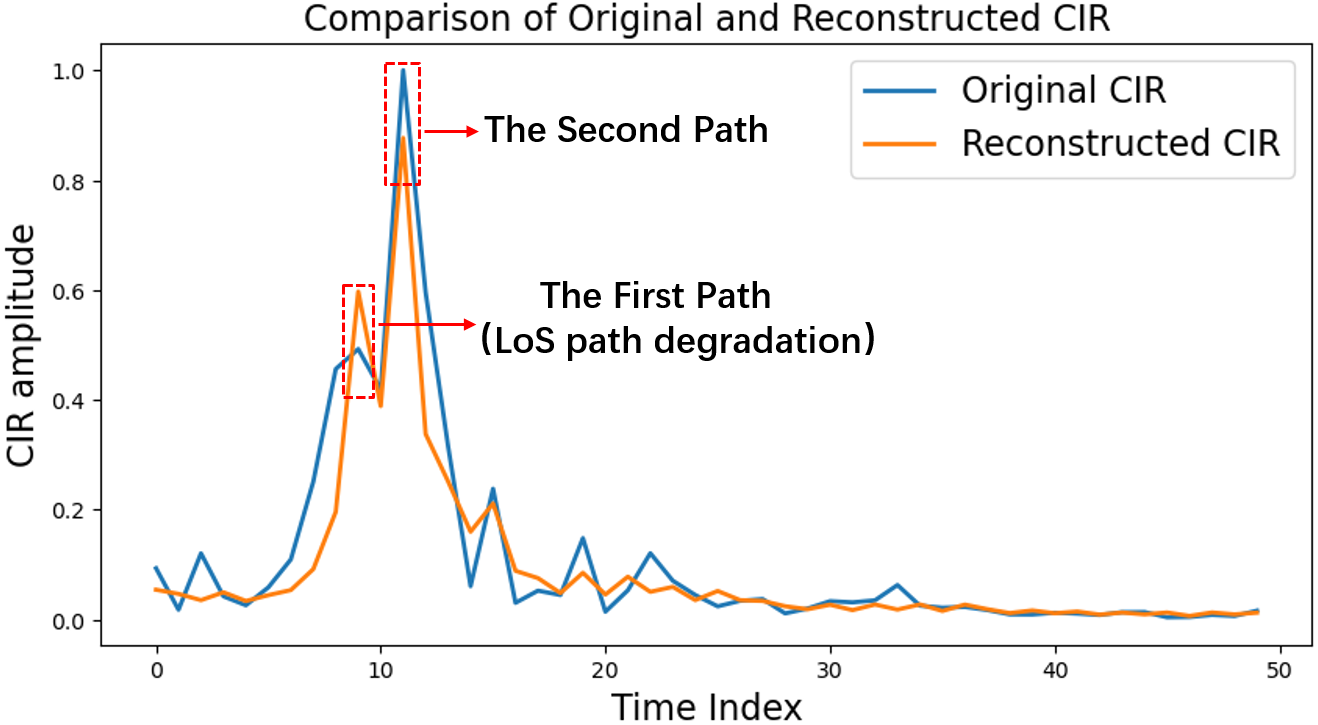}}
\caption{(a) Comparison of the original and generated CIR under the LOS condition. (b) Comparison of the original and generated CIR under the NLOS condition (LOS path degradation).}~\label{fig:generateCIR}
\end{figure*}

Table~\ref{tab:accuracy} and Fig.~\ref{fig:cdf} (a) (b) display the localization accuracy of various methods under the aforementioned two scenarios. Specifically, our proposed GenMetaLoc outperforms others, achieving 80th percentile errors of 1.65 m in LOS and 1.98 m in NLOS environments. The relatively lower performance in the lab is due to the complex environment with obstacles, which causes stronger multipath effects. By contrast, the method using only CSI amplitude, though utilizing the same framework as GenMetaLoc, shows lower performance in both scenarios, particularly in the challenging NLOS environment. This highlights the effectiveness of our data processing methods that incorporate both CSI amplitude and CSI phase. Additionally, MetaLoc using CIR exhibits lower performance than GenMetaLoc due to the absence of environmental information, but it still outperforms the CSI amplitude-only method. On the other hand, the competing method ConFi achieves the second-lowest localization error in the LOS scenario; however, it requires a substantial amount of data for training in new environments~\cite{10274764}. RI with random initialization from the generative network exhibits slower convergence and lower localization performance as shown in Fig.~\ref{fig:meta-test-loss}, which highlights the effectiveness of the meta-parameters trained using the learning-to-learn framework. Finally, KNN shows the worst performance, primarily due to its limited robustness to outliers introduced by changed environmental conditions. 
\begin{figure*}
\centering
\subfigure[]{\includegraphics[scale=0.4]{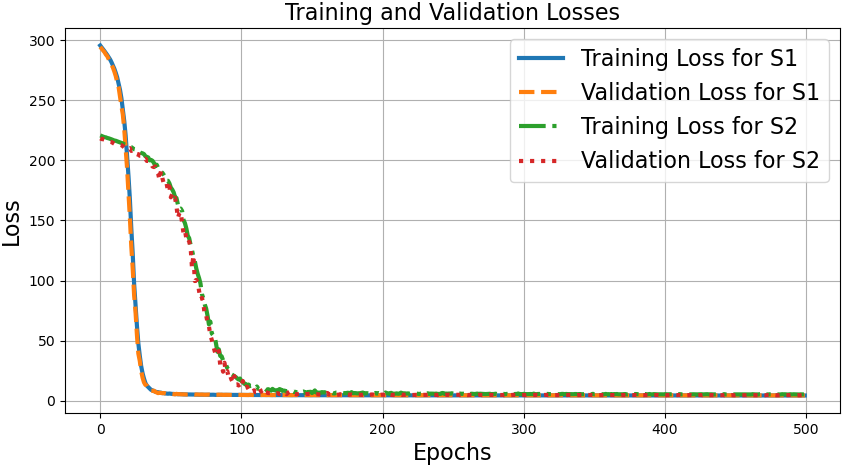}}
\subfigure[]{\includegraphics[scale=0.385]
{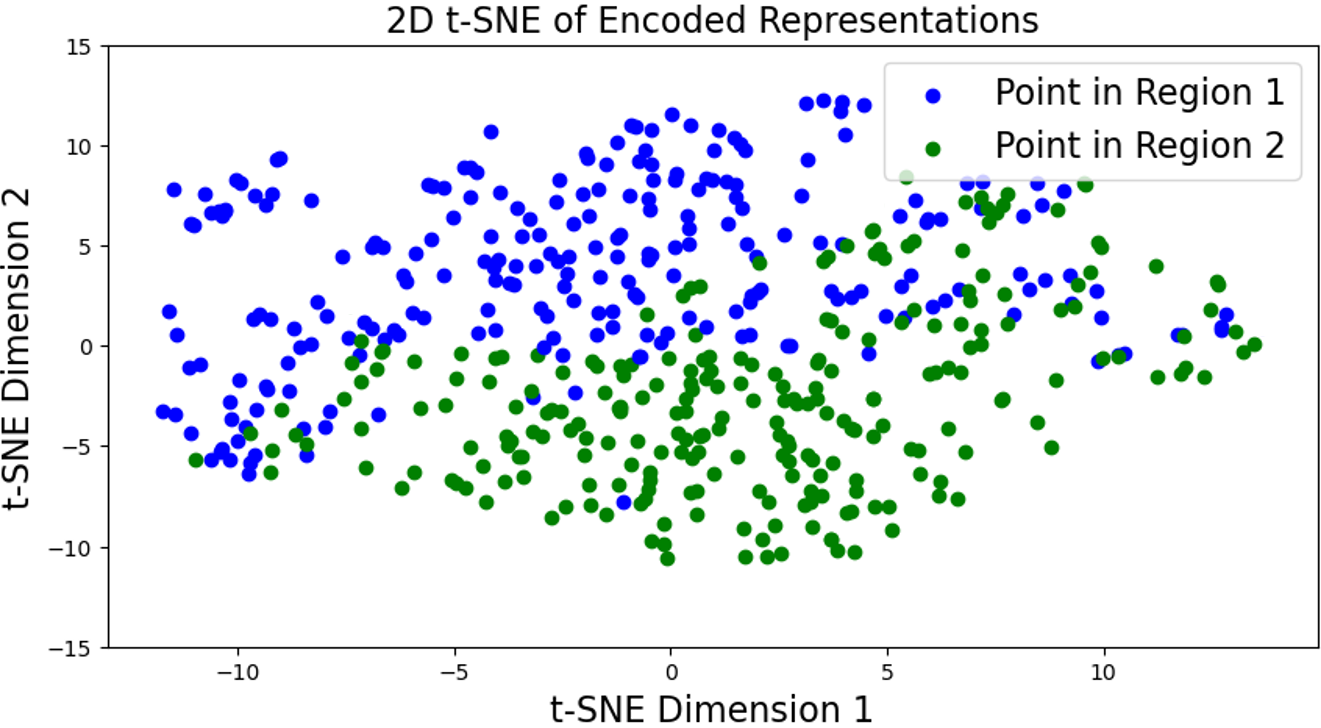}}
\caption{(a) Training and validation losses of the CIR autoencoder under the S1 and S2 scenarios. (b) Two-dimensional t-SNE representation of the latent space, dividing the whole experimental area into two regions: the first 250 points (Region 1) in blue and the last 250 points (Region 2) in green.}
\label{fig:vae}
\end{figure*}

\subsection{Further Analysis}
To further analyze the superior performance of GenMetaLoc compared to other methods, we examine the CIR profiles generated by our newly introduced diffusion model, as shown in Fig.~\ref{fig:generateCIR} (a) and (b). Specifically, subfigure (a) illustrates the presence of a LOS path between the transmission pair, characterized by the first multipath component exhibiting the highest amplitude. In contrast, subfigure (b) depicts an NLOS scenario, where the first path exhibits a weaker amplitude than the second path, indicating a degraded LOS multipath component. In both LOS and NLOS conditions, the generated CIRs effectively preserve the main path information and smooth out the noise, demonstrating the model's capability to capture the underlying data structure from different propagation conditions.

Figure~\ref{fig:vae} presents the performance of the CIR autoencoder from different perspectives. Specifically, subfigure (a) illustrates the training and validation losses of the autoencoder for the S1 and S2 scenarios. All loss curves achieve convergence, indicating that the model has been well-trained without overfitting. Moreover, the loss curves for S1 converge more rapidly than those for S2, as S1 is a LOS-dominant environment that enables the collection of higher quality training data. Subfigure (b) illustrates the two-dimensional t-SNE visualization of the latent space generated by the encoder, where the experimental area is divided into two regions: the first 200 points (Region 1) are shown in blue, and the last 200 points (Region 2) are shown in green. We can see that the latent representation effectively separates points from different physical regions. Specifically, points from Region 1 are distributed in the top area of the latent space, while points from Region 2 are concentrated in the bottom area.  The clear separation of points in the latent space demonstrates that the autoencoder has effectively learned location-specific features, enabling efficient compression of CIR data and subsequent generation of high-quality synthetic fingerprints for enhancing the localization process.
\section{Conclusion}
This paper presents GenMetaLoc, the first meta-learning framework utilized in the generation of dense fingerprint databases from an environment-aware perspective. In terms of model convergence, GenMetaLoc shows faster convergence with minimal data samples, outperforming methods that rely on random initialization. With regard to localization accuracy, compared to MetaLoc that does not incorporate environmental information, GenMetaLoc achieves performance improvements of 14.06\% in the LOS environment and 15.74\% in the NLOS environment. In terms of data processing, GenMetaLoc outperforms methods that rely solely on CSI amplitude, yielding improvements of approximately 19.51\% in the LOS environment and 21.74\% in the NLOS environment. Our work leverages the sensing capabilities of existing commercial WiFi devices and mobile phones, contributing to the development of ISAC systems, and facilitating efficient WiFi localization and sensing for future cost-effective communication networks.
\bibliographystyle{IEEEtran}
\bibliography{Refs.bib}

\end{document}